\documentclass[12pt, draftclsnofoot,onecolumn]{IEEEtran}
\usepackage{amsmath}
\usepackage{bm, amssymb}
\usepackage{lipsum}
\usepackage{graphicx}
\usepackage{algorithm}
\usepackage{algpseudocode}
\usepackage{amsmath}
\usepackage{graphics}
\usepackage{epsfig}
\usepackage{subfigure}
\usepackage{booktabs}
\usepackage{multirow}
\usepackage{makecell}
\usepackage{color}

\newtheorem{theorem}{Theorem}
\newtheorem{proposition}{Proposition}
\newtheorem{lemma}{Lemma}
\newtheorem{corollary}{Corollary}
\newtheorem{definition}{Definition}
\newtheorem{remark}{Remark}

\ifCLASSINFOpdf
\else
\fi
\begin{document}
%
\title{An Entropy-based Proof of Threshold Saturation for Nonbinary SC-LDPC\\ Ensembles on the BEC}
%
%
%

\author{Zhonghao~Zhang,~
        Mengnan~Xu,~
        Chongbin~Xu,~
        Dan~Zeng,~
        and~Zhichao~Sheng~

\thanks{Manuscript received ~~~ ~~, 2020; revised ~~~ ~~, 2020; accepted ~~~ ~~, ~~~~. Date of publication ~~~ ~~, ~~~~; date of current version ~~~ ~~, ~~~~. The associate editor coordinating the review of this paper and approving it for publication was         ~~~~~~~~~.}

}

%
%

\markboth{IEEE Transactions on ,~Vol.~X, No.~X, X~20XX}%
{Zhang \MakeLowercase{\textit{et al.}}: An Entropy-based Proof of Threshold Saturation for Nonbinary SC-LDPC Ensembles on the BEC}
%



\maketitle

\begin{abstract}
In this paper we are concerned with the asymptotic analysis of nonbinary spatially-coupled low-density parity-check (SC-LDPC) ensembles defined over GL$\left(2^{m}\right)$ (the general linear group of degree $m$ over GF$\left(2\right)$).
Our purpose is to prove threshold saturation when the transmission takes place on the binary erasure channel (BEC).
To this end, we establish the duality rule for entropy for nonbinary variable-node (VN) and check-node (CN) convolutional operators to
accommodate the nonbinary density evolution (DE) analysis.
Based on this, we construct the explicit forms of the potential functions for uncoupled and coupled DE recursions.
In addition, we show that these functions exhibit similar monotonicity properties as those for binary LDPC and SC-LDPC ensembles over general binary memoryless symmetric (BMS) channels.
This leads to the threshold saturation theorem and its converse for nonbinary SC-LDPC ensembles on the BEC, following the proof technique developed by S. Kumar \textit{et al}.
\end{abstract}


%
\IEEEpeerreviewmaketitle

\section{Introduction}
%
%
%
%

\IEEEPARstart{S}{patial} coupling has been recognized as an effective way of improving the performance of low-density parity-check (LDPC) codes.
This concept was first introduced in \cite{5695130}, but its underlying idea can be traced back to the benchmark work by Zigangirov \cite{782171} for the design of LDPC codes with convolutional structures.
The resultant codes, termed as spatially-coupled LDPC (SC-LDPC) codes, are found to have better error correction capability than the uncoupled ones in terms of decoding threshold  \cite{1362891}\cite{5571910}.
This finding motivates the applications of the underlying principle behind SC-LDPC codes to a wide variety of communication systems with much success.
See \cite{6655077}\cite{6990513}\cite{7084636} for coded modulation systems, \cite{6034031}\cite{6364433}\cite{6387167} for inter-symbol interference channels, and \cite{6034088}\cite{6120387}\cite{7031889} for multiple access channels.

From a design point of view, it is of particular importance to predict the asymptotic performance gain introduced by SC-LDPC codes compared with standard LDPC block codes.
This can be done by calculating the belief propagation (BP) threshold of SC-LDPC codes based on the coupled DE algorithm, but at a cost of high complexity.
A more efficient way is to prove the existence of the \textit{threshold saturation} effect.
For example, for a binary regular SC-LDPC code, it has been shown that the BP threshold \textit{saturates} to the maximum-a-posteriori (MAP) threshold of its underlying uncoupled LDPC codes on the BEC \cite{5695130} and general binary memoryless symmetric (BMS) channels \cite{6589171}.
As the MAP threshold of the binary \textit{regular} LDPC code can be (tightly) calculated based on the generalized extrinsic information transfer (GEXIT) chart \cite{Richardson:2008:MCT:1795974}, this theoretic result provides a simple guidance to predetermine the asymptotic BP threshold of a regular SC-LDPC code, avoiding the need of the coupled DE algorithm.

{For general SC-LDPC coded systems characterized by scalar recursions (e.g., the DE recursion for a binary \textit{irregular} SC-LDPC code on the BEC)}, Yedla \textit{et al.} introduced a technique based on potential functions for the proof of threshold saturation \cite{6325197}\cite{6887298}.
The underlying idea behind this technique is to construct real-valued potential functions by taking an integral of scalar DE recursions (e.g., the areas under the transfer curves in the EXIT chart).
By doing this, Yedla \textit{et al.} proceeded the analysis of the DE fixed points by investigating the stationary points of the potential functions.
They proved that, the threshold of a scalar coupled DE recursion asymptotically coincides with the potential threshold of the uncoupled DE recursion defined by the vanishing of the so-called energy gap (a local minimum of the underlying potential function).
This technique can be directly applied to binary irregular SC-LDPC codes on the BEC, proving the existence of the threshold saturation effect in this scenario.

The work by \cite{6325197}\cite{6887298} was proposed for scalar recursions.
The main difficulty of its extension to nonscalar recursions is how to construct potential functions by taking an integral over the space of density functions.
{For SC-LDPC codes on the BMS channels, S. Kumar \textit{et al} circumvented this difficulty by} specializing the replica-symmetric (RS) free entropy functional to LDPC ensembles and derived the potential functions based on entropies \cite{6912949}.
It turns out that these functions are the negative of the RS free entropies associated with the code ensembles.
Their analysis shed a light on the invaluable role of the duality rule for entropy \cite{Richardson:2008:MCT:1795974} in the construction of potential functions.
This rule reveals an entropy conservation relation involving the variable-node (VN) and the check-node (CN) convolutional operators \cite{measson2006conservation}, establishing the bridge between the DE fixed points and the stationary points of potential functions.
Following the idea by S. Kumar \textit{et al.}, we are able to extend the entropy-based proof technique to binary irregular SC-LDPC ensembles on general BMS channels.

The performance gain introduced by employing \textit{nonbinary} SC-LDPC codes has been numerically observed \cite{5706907}\cite{6874959}\cite{7024893}.
It arises a natural question whether the threshold saturation effect also exists in such scenarios.
Motivated by this, the authors in \cite{7430313} studied nonbinary SC-LDPC codes defined over the general linear group when the transmission takes place on the BEC.
They concluded that, to apply the proof technique by Yedla \textit{et al}, one should first identify the existence of the potential functions for the {nonscalar} DE recursions.
For this reason, the authors developed a constructive criterion that is applicable to general vector spatially-coupled recursions defined over general multivariate polynomials.
Although the authors conjectured that potential functions always exist, it seems not an easy task to construct these functions except for some special cases (see Table II therein).

In this paper, we focus on the asymptotic performance of nonbinary SC-LDPC ensembles defined over the general linear group GL$\left(2^{m}\right)$ and prove that the threshold saturation effect indeed occurs for transmission on the BEC.
Our work is a nonstraightforward extension of \cite{7430313} and \cite{6912949}.
Our contribution is three-fold.

\begin{itemize}
\item First of all, we establish the duality rule for entropy for nonbinary DE recursions on the BEC.
    As in the binary case mentioned above, this rule also reveals a conservation relation between the input and the output entropy of \textit{nonbinary} VN and CN convolutional operators and is the key step towards constructing potential functions in the proof of threshold saturation.
\item Secondly, we propose the explicit forms of nonbinary potential functions similar to those in \cite{6912949} derived for binary SC-LDPC ensembles over BMS channels.
    This proves the conjecture proposed in \cite{7430313} for all code degree distributions and $m$.
    We further show that these potential functions exhibit similar monotonicity properties including the partial order preservation properties.
    This finding implies that it is possible to develop the threshold saturation theorem and its converse for nonbinary SC-LDPC ensembles on the BEC, following the idea by S. Kumar \textit{et al} \cite{6912949}.
\item Finally, we modify the definition of the energy gap that is used to calculate the potential threshold of the underlying LDPC ensemble. In specific, the energy gap in \cite{6912949} is defined based on the infimum over the complementary subset of the basin of attraction to the trivial DE fixed point, while in our work we restrict the complementary subset to the set of nontrivial underlying DE fixed points (see Definition \ref{DEF: Energy Gap}).
\end{itemize}

The remainder of the paper is organized as follows.
In Section \ref{Preliminaries}, we define nonbinary LDPC and SC-LDPC ensembles concerned in this paper and briefly discuss the form of the density in the nonbinary DE analysis.
In Section \ref{Duality Rule for Entropy and Partial Ordering}, we review the definitions of the entropy function and the VN and CN convolutional operators.
We establish and prove several important identities and properties including the duality rule for entropy and the partial order preservation properties.
In Section \ref{Potential Analysis and Threshold Saturation}, we construct potential functions for nonbinary uncoupled and coupled DE recursions.
The monotonicity properties of these functions are also proposed and proved based on the theoretic results in Section \ref{Duality Rule for Entropy and Partial Ordering}.
We establish the threshold saturation theorem and its converse at the end of Section \ref{Potential Analysis and Threshold Saturation}.
Finally, Section \ref{Conclusion} concludes the whole paper.

\subsection{Notations}

We use $\mathbb{R}$ to represent the set of all real numbers and define $\mathbb{Z} = \left\{0, 1, 2, \ldots \right\}$ and $\mathbb{Z}_{+} = \mathbb{Z}\backslash \left\{0\right\}$.
For any $m \in \mathbb{Z}_{+}$, we define $\mathbb{M} = \left\{0, 1, \ldots, m\right\}$.
The two integers $N$ and $w$ denote the coupling length and the coupling width for an SC-LDPC ensemble, respectively.
By defining $N_{w} = N + w - 1$, we introduce $\mathbb{N}_{\text{v}} = \left\{0, 1, \ldots, N - 1\right\}$ and $\mathbb{N}_{\text{c}} = \left\{0, 1, \ldots, N_{w} - 1\right\}$ to denote the positions of VNs and CNs, respectively.
Further, define $N^{\text{mid}}_{w} = \lfloor \left(N + w - 1\right)/2 \rfloor$ where $\lfloor x \rfloor$ represents the maximum integer less than or equal to $ x \in \mathbb{R}$.

\section{Preliminaries}
\label{Preliminaries}
\subsection{LDPC and SC-LDPC Ensembles Defined Over GL$\left(2^{m}\right)$}

Denote by LDPC$\left(\lambda, \rho, {m}\right)$ the nonbinary LDPC ensemble defined over the general linear group GL$\left(2^{m}\right)$.
Here we omit the codeword length for notational brevity, since in this paper we always restrict ourselves to the limit where the codeword length trends to infinity.
Following the standard notational convention, we use $\lambda\left(x\right) = \sum_{i}{\lambda_{i}x^{i-1}}$ and $\rho\left(x\right) = \sum_{j}{\rho_{j}x^{j-1}}$ to denote the edge-perspective degree distributions of VNs and CNs, respectively, with nonnegative coefficients $\lambda_i$ and $\rho_j$ satisfying $\lambda\left(1\right) = \rho\left(1\right) = 1$.
We also adopt node-perspective degree distributions denoted as $L\left(x\right) = \sum_{i}{L_{i}x^{i}}$ and $R\left(x\right) = \sum_{j}{R_{j}x^{j}}$, the coefficients of which are determined by \cite{Richardson:2008:MCT:1795974}
\begin{equation}
\label{Definition: Relationship Between EPDD and NPDD}
L_{i} = \frac{{\lambda_{i}}/{i}} {\sum_{k}{\lambda_{k}}/{k}}, \quad R_{j} = \frac{{\rho_{j}}/{j}}{\sum_{k}{\rho_{k}}/{k}}.
\end{equation}

A nonbinary LDPC code selected from LDPC$\left(\lambda, \rho, {m}\right)$ can be described in the form of a bipartite graph termed the Tanner graph.
Each VN $i$ in the Tanner graph corresponds to a coded symbol defined over GF$\left(2^{m}\right)$.
When the transmission takes place on the BEC, it is convenient to write the coded symbol in the form of a binary column vector of $m$ bits, i.e. $\textbf{x}_{i} = \left(x_{i, 1}, x_{i, 2}, \ldots, x_{i, m} \right)^{\text{T}}$ with $x_{i, k} \in \left\{0, 1\right\}, \forall k \in \mathbb{M} \backslash \left\{0\right\}$.
With this notation, we can represent the coding constraint imposed by each CN $a$ as follows
\begin{equation}
\label{Definition: Linear Constraint Imposed by a CN}
\sum_{i \in \partial{a}} \textbf{W}_{i, a} \textbf{x}_{i} = \textbf{0}
\end{equation}
where $\textbf{0}$ denotes the zero vector of length $m$, $\partial a$ the subset of VNs connected to CN $a$, and $\textbf{W}_{i, a}$, a binary $m$-by-$m$ invertible matrix uniformly selected from GL$\left(2^{m}\right)$ at random, is the label of the edge from VN $i$ to CN $a$ in the Tanner graph.

\begin{figure*}[ht]
\begin{center}
\scalebox{1}{\includegraphics[height=3.4cm ,width=19.90cm, angle=0, scale=0.8]{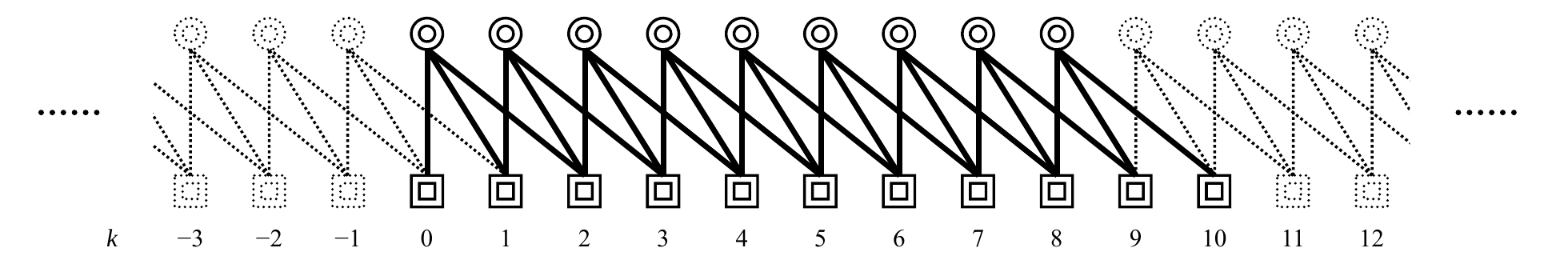}}
\end{center}
\caption{ The Tanner graph for SC-LDPC$\left(\lambda, \rho, 9, 3, m\right)$, where each square (resp., circle) represents a collection of multiple CNs (resp., VNs) of the underlying LDPC ensemble located at that position. The dashed squares, circles and edges are removed in the termination procedure.
}
\label{Figure TG for SC-LDPC}
\end{figure*}

We also consider the nonbinary SC-LDPC ensemble over GL$\left(2^{m}\right)$ denoted as SC-LDPC$\left(\lambda, \rho, \right.$ $\left.N, w, {m}\right)$ in this paper.
Such an ensemble can be constructed from the graphic perspective as follows.
As illustrated in Fig. \ref{Figure TG for SC-LDPC}, we first place the Tanner graphs of LDPC$\left(\lambda, \rho, {m}\right)$ along a chain, the positions of which are indexed by an integer $k$.
Next, at each position $k$, the outgoing edges of the VNs are uniformly and randomly divided into $w$ groups, being reconnected to those CNs at positions $\left\{k, k+1, \ldots, k + w - 1\right\}$.
Likewise, the CNs at each position $k$ are also uniformly and randomly connected to the VNs at positions $\left\{k - w + 1, k - w + 2, \ldots, k\right\}$.
After that, we terminate the coupling chain by removing the VNs at positions $\left\{\ldots, -2, -1\right\} \cup \left\{N, N + 1, \ldots\right\}$ and their outgoing edges.
As a result, all CNs with degree less than two become invalid and thus are also removed from the coupling chain.
The resultant graph is referred to as the Tanner graph of SC-LDPC$\left(\lambda, \rho, N, w, {m}\right)$.
The termination procedure will reduce the degrees of some CNs at the two ends of the coupling chain.
A coding rate loss is introduced, but it will vanish as $N \to \infty$ (while keeping $w$ fixed).
More importantly, the termination procedure leads to a phenomenon termed decoding wave propagation in the BP decoding algorithm, which is the fundamental mechanism behind threshold saturation.

One may equivalently define SC-LDPC$\left(\lambda, \rho, N, w, {m}\right)$ from the parity-check matrix perspective. See \cite{7430313} for details.

\subsection{Densities of Messages in BP Decoding}
\label{SubSec: Densities of Messages in BP Decoding}
In the DE analysis, we are interested in tracking the distributions of messages exchanged in the BP decoding algorithm.
These distributions are referred to as the densities.
In general, density tracking is difficult for nonbinary LDPC ensembles since the decoding performance may depend on the transmitted codeword with $2^{m}$ possible values for each coded symbol $\textbf{x}_{i}$.
Fortunately, in the case where the transmission takes place on the BEC and the edge labels are defined over GL$\left(2^{m}\right)$, the form of the density can be simplified.
First of all, thanks to the symmetry of the BEC, the BP decoding performance does not depend on the specific transmitted codeword, therefore we can assume that the all-zero codeword is transmitted \cite{Richardson:2008:MCT:1795974}.
Under this assumption, the a posteriori probability mass function (PMF) of $\textbf{x}_{i}$ is equiprobable over a subspace $\mathcal{S}$ of the $m$-dimensional binary vector space \cite{1561993}.
Consider an example where $m = 3$, $\textbf{x}_{i} = \left(0, 0, 0\right)^{\text{T}}$ and $\textbf{y}_{i} = \left(0, ?, ?\right)^{\text{T}}$ with $\textbf{y}_{i}$ being the channel observation containing $k = 2$ erased bits ``$?$''.
In this example, the a posteriori PMF is given by $p\left(\textbf{x}_{i} | \textbf{y}_{i}\right) = {1}/{2^{k}} = 0.25$ if $\textbf{x}_{i}$ takes values from the subspace $\mathcal{S} = \left\{\left(0, 0, 0\right)^{\text{T}}, \left(0, 0, 1\right)^{\text{T}}, \left(0, 1, 0\right)^{\text{T}}, \left(0, 1, 1\right)^{\text{T}}\right\}$ of dimension $k = 2$, and $p\left(\textbf{x}_{i} | \textbf{y}_{i}\right) = 0$ otherwise.
Secondly, it can be shown that the subspace dimension does not change when a message is passed along an edge in the BP decoding algorithm.
To see this, notice that if the a posteriori PMF of $\textbf{x}_{i}$ is equiprobable over $\mathcal{S}$, then the a posteriori PMF of $\textbf{W}_{i, a}\textbf{x}_{i}$ is equiprobable over $\mathcal{S}' = \left\{\textbf{x}' | \textbf{x}' = \textbf{W}_{i, a}\textbf{x}, \forall \textbf{x} \in \mathcal{S}\right\}$.
Obviously, the dimensions of $\mathcal{S}'$ and $\mathcal{S}$ are identical due to the fact that the binary matrix $\textbf{W}_{i, a}$ is invertible.
Therefore, it is sufficient to keep track the subspace dimensions instead of the a posteriori PMFs of coded symbols \cite{1561993}.

For the above reason, in this paper, our discussions are based on the density with the following form as in \cite{7430313}\cite{1561993}.

\begin{definition}
\label{DEF: Densities}
The density of a message in the BP decoding algorithm for LDPC$\left(\lambda, \rho, {m}\right)$ and SC-LDPC$\left(\lambda, \rho, N, w, {m}\right)$ on the BEC is defined as the probability vector of length $m+1$, the $k$-th entry of which is the probability that the a posteriori PMF corresponding to the message is equiprobable over a subspace of dimension $k, \forall k \in \mathbb{M}$. In what follows, the set of all such densities will be denoted as $\mathcal{X}$, i.e.,
\begin{equation}
\label{Definition: Density Set}
\mathcal{X} = \left\{ \mathtt{a} = \left(a_0, a_1, \ldots, a_m\right) \left| \sum_{k = 0}^{m}a_k = 1, a_k \ge 0, k \in \mathbb{M} \right. \right\}.
\end{equation}
For notational brevity, we will also use $\left[\mathtt{a}\right]_{k}$ to represent the $k$-th entry of $\mathtt{a}$, $\forall k \in \mathbb{M}$,
\end{definition}

There are two extremal densities in $\mathcal{X}$, one of which is $\mathtt{\Delta}_{m} = \left(0, 0, \ldots, 0, 1\right)$ corresponding to the case where the message offers no information about the coded symbol, and the other is $\mathtt{\Delta}_{0} = \left(1, 0, \ldots, 0, 0\right)$ corresponding to the error-free case where the coded symbol can be recovered from the message perfectly.
Further, we will use $\mathtt{\Delta}_{k}$ to denote the density with the $k$-th entry being $1$ and others being $0$, $\forall k \in \mathbb{M}$.

\section{Duality Rule for Entropy and\\ Partial Ordering}
\label{Duality Rule for Entropy and Partial Ordering}
\subsection{The Duality Rule for Entropy}

In this subsection, we will establish the duality rule for entropy for nonbinary LDPC and SC-LDPC ensembles on the BEC.
To this end, we first present and review the definitions of the entropy function and the basic VN and CN operators.

\begin{definition}
\label{DEF: Entropy Function}
For any $\mathtt{a} \in \mathcal{X}$, the entropy function of $\mathtt{a}$ is defined as
\begin{equation}
H\left(\mathtt{a}\right) = \sum_{k = 1}^{m} k a_{k}.
\end{equation}
\end{definition}

\begin{remark}
\label{RMK: Discussion On H}
The entropy function $H\left(\mathtt{a}\right)$ can be regarded as a measure of the average uncertainty of a message, the distribution of which can be determined by $\mathtt{a}$.
As discussed in Subsection \ref{SubSec: Densities of Messages in BP Decoding}, the a posteriori PMF of a message is always equiprobable over a subspace $\mathcal{S}$.
Let $k$ be the dimension of $\mathcal{S}$.
Since there are $2^k$ elements in $\mathcal{S}$ with equal probability, the uncertainty of this message is $k$ bits.
Therefore, if we treat $k$ as a random variable with $\mathtt{a}$ being the distribution, then the average uncertainty of the message is given by $\sum_{k = 1}^{m} k a_{k}$ bits.
\end{remark}

In this paper, we adopt the notions $\boxdot$ and $\boxtimes$ introduced in \cite{1561993} for the VN and CN convolutional operators.

\begin{definition}
\label{DEF: Basic Operators at VNs and CNs}
For any $\mathtt{a}, \mathtt{b} \in \mathcal{X}$, $\mathtt{a} \boxdot \mathtt{b}$ and $\mathtt{a} \boxtimes \mathtt{b}$ are two densities, the $k$-th entries of which are respectively given by
\begin{equation}
\label{Definition: Basic Operators at VNs and CNs}
\left[\mathtt{a} \boxdot \mathtt{b}\right]_k = \sum_{i=0}^{m} \sum_{j=0}^{m} a_{i}V_{i, j, k}^{m}b_{j}, \quad \left[\mathtt{a} \boxtimes \mathtt{b}\right]_k = \sum_{i=0}^{m} \sum_{j=0}^{m} a_{i}C_{i, j, k}^{m}b_{j}
\end{equation}
$\forall k \in \mathbb{M}$.
Here, the coefficients $V_{i, j, k}^{m}$ and $C_{i, j, k}^{m}$ are respectively given by
\begin{equation}
\label{Definition: Vijk and Cijk}
V_{i, j, k}^{m} = \frac{2^{(i - k)(j - k)} \genfrac{[}{]}{0pt}{0}{i}{k} \genfrac{[}{]}{0pt}{0}{m - i}{j - k}} {\genfrac{[}{]}{0pt}{0}{m}{j}}, \quad C_{i, j, k}^{m} = \frac{2^{(k - i)(k - j)} \genfrac{[}{]}{0pt}{0}{m - i}{m - k} \genfrac{[}{]}{0pt}{0}{i}{k - j}} {\genfrac{[}{]}{0pt}{0}{m}{m - j}}
\end{equation}
with $\genfrac{[}{]}{0pt}{1}{m}{k}$ being the Gaussian binomial coefficient defined as follows
\begin{equation}
\label{Definition: Gmk}
\genfrac{[}{]}{0pt}{0}{m}{k} =
\begin{cases}
1,  & k = 0 \text{ or } k = m \\
\prod_{l = 0}^{k - 1}\frac{2^{m}-2^{l}}{2^{k}-2^{l}}, &0 < k < m \\
0, &\text{otherwise}.
\end{cases}
\end{equation}

In addition, we define $\mathtt{a}^{\boxdot n} = \underbrace{\mathtt{a} \boxdot \mathtt{a} \boxdot \ldots \boxdot \mathtt{a}}_{n \text{ terms } \mathtt{a}}$ and $\mathtt{a}^{\boxtimes n} = \underbrace{\mathtt{a} \boxtimes \mathtt{a} \boxtimes \ldots \boxtimes \mathtt{a}}_{n \text{ terms } \mathtt{a}}$ for $n \in \mathbb{Z}_{+}$, and we use the convention that $\mathtt{a}^{\boxdot 0} = \mathtt{\Delta}_{m}$ and $\mathtt{a}^{\boxtimes 0} = \mathtt{\Delta}_{0}$ if $\mathtt{a} \in \mathcal{X}$.
\end{definition}

For notational convenience, in the sequel, we will use $*$ to denote either $\boxdot$ or $\boxtimes$.
In Appendix \ref{APP: Three Laws of Basic Operators}, we will prove the commutative, distributive and associative laws of $*$ and apply them to the derivative analysis of the entropy function.

\begin{remark}
In the sequel, we will compute the difference between the entropies of two densities involving the convolutional operator $*$, e.g., $H\left(\mathtt{a} * \mathtt{c}\right) - H\left(\mathtt{b} * \mathtt{c}\right), \forall \mathtt{a}, \mathtt{b}, \mathtt{c} \in \mathcal{X}$.
For notational convenience, we will extend Definitions \ref{DEF: Entropy Function} and \ref{DEF: Basic Operators at VNs and CNs} to all real-valued vectors of length $m + 1$ (not necessarily the probability vectors).
By doing this, we can rewrite $H\left(\mathtt{a} * \mathtt{c}\right) - H\left(\mathtt{b} * \mathtt{c}\right)$ as $H\left(\left(\mathtt{a} -\mathtt{b}\right) * \mathtt{c}\right)$.
\end{remark}

We are now ready for the duality rule for entropy for nonbinary LDPC and SC-LDPC ensembles on the BEC.
\begin{lemma}
\label{LEM: Duality Rule for Entropy}
For any $\mathtt{a}, \mathtt{b} \in \mathcal{X}$,
\begin{equation}
\label{Lemma: Duality Rule for Entropy}
H\left(\mathtt{a}\right) + H\left(\mathtt{b}\right) = H\left(\mathtt{a}\boxdot\mathtt{b}\right) + H\left(\mathtt{a}\boxtimes\mathtt{b}\right).
\end{equation}
\end{lemma}

\begin{IEEEproof}
By Definition \ref{DEF: Entropy Function},
\begin{equation}
\label{Proof: kV + kC = i + j}
H\left(\mathtt{a}\boxdot\mathtt{b}\right) + H\left(\mathtt{a}\boxtimes\mathtt{b}\right) - H\left(\mathtt{a}\right) - H\left(\mathtt{b}\right) = \sum_{i=0}^{m} \sum_{j=0}^{m} a_{i} b_{j} \left[\sum_{k = 1}^{m} k\left(V_{i, j, k}^{m} + C_{i, j, k}^{m}\right) - \left(i + j\right)\right].
\end{equation}

Therefore, it is suffice to show that, for any $i, j \in \mathbb{M}$,
\begin{equation}
\label{Lemma: Sum of kV+kC over k}
\sum_{k = 1}^{m} k\left(V_{i, j, k}^{m} + C_{i, j, k}^{m}\right) = i + j.
\end{equation}

Although (\ref{Lemma: Sum of kV+kC over k}) can be verified for small values of $m$, how to prove it for all $m \in \mathbb{Z}_{+}$ is the most difficult step in the proof of Lemma \ref{LEM: Duality Rule for Entropy}.
One may consider the method by induction on $m$.
However, such an idea is perhaps not feasible since the relation between $V_{i, j, k}^{m}$ and $V_{i, j, k}^{m+1}$ is quite involved.
To circumvent this difficulty, we construct two bivariate functions (see (\ref{Proof: Procedure to f(x,y) 2}) and (\ref{Proof: Intermediate Results for kC}) below), and by taking their partial derivatives we will obtain two polynomials whose coefficients are related to the left-hand side of (\ref{Lemma: Sum of kV+kC over k}).
This will lead to the desired result.

Now we proceed the proof of Lemma \ref{LEM: Duality Rule for Entropy} with the following Gauss's binomial formula \cite{kac2001quantum},
\begin{equation}
\label{Basic Identity Prod = Sum}
\prod_{\alpha = 1}^{m}\left(1 + 2^{\alpha - 1}x\right) = \sum_{j = 0}^{m}2^{\frac{1}{2}j\left(j-1\right)}\genfrac{[}{]}{0pt}{0}{m}{j} x^{j}, \quad x \in \mathbb{R}.
\end{equation}

For $x, y > 0$, define $f\left(x, y\right)$ as follows
\begin{equation}
\label{Definition: f(x,y)}
f\left(x, y\right) = \prod_{\alpha = 1}^{i}\left(1 + 2^{\alpha - 1}x\right)  \prod_{\beta = 1}^{m - i}\left(1 + 2^{\beta + i - 1}y\right).
\end{equation}

By applying (\ref{Basic Identity Prod = Sum}) to (\ref{Definition: f(x,y)}), we obtain
\begin{align}
f\left(x, y\right) =& \sum_{l_{1} = 0}^{i} 2^{\frac{1}{2}l_{1}\left(l_{1}-1\right)}\genfrac{[}{]}{0pt}{0}{i}{l_{1}} x^{l_{1}} \sum_{l_{2} = 0}^{m-i} 2^{\frac{1}{2}l_{2}\left(l_{2}-1\right)}\genfrac{[}{]}{0pt}{0}{m-i}{l_{2}} \left(2^{i}y\right)^{l_{2}} \\
\stackrel{(a)}{=}& \sum_{k = 0}^{i} \sum_{j = k}^{m -i + k} 2^{\frac{1}{2}j\left(j-1\right) + \left(i-k\right)\left(j-k\right)}\genfrac{[}{]}{0pt}{0}{i}{k} \genfrac{[}{]}{0pt}{0}{m-i}{j-k} x^{k} y^{j-k} \\
\label{Relationship: f(x,y) and V}
=& \sum_{j = 0}^{m}  2^{\frac{1}{2}j\left(j-1\right)} \genfrac{[}{]}{0pt}{0}{m}{j} \sum_{k = 0}^{m} V_{i, j, k}^{m} x^{k} y^{j-k}
\end{align}
where (a) is obtained by replacing $l_{1}$ and $l_{2}$ with $k$ and $j - k$, respectively.


Now we take the partial derivative of $f\left(x, y\right)$ in (\ref{Definition: f(x,y)}) with respect to $x$, then multiply the result by $x$, and finally replace $y$ with $x$. This leads to the following result
\begin{equation}
\label{Proof: Procedure to f(x,y) 1}
\left.\left[x \frac{\partial}{\partial x}f\left(x, y\right)\right]\right|_{y = x} = \left(\sum_{\alpha = 1}^{i}\frac{2^{\alpha - 1}x}{1 + 2^{\alpha - 1}x}\right) \prod_{\beta = 1}^{m}\left(1 + 2^{\beta - 1}x\right).
\end{equation}

Applying the same procedure to (\ref{Relationship: f(x,y) and V}) yields
\begin{equation}
\label{Proof: Procedure to f(x,y) 2}
\left.\left[x \frac{\partial}{\partial x}f\left(x, y\right)\right]\right|_{y = x} = \sum_{j = 0}^{m}  2^{\frac{1}{2}j\left(j-1\right)} \genfrac{[}{]}{0pt}{0}{m}{j} x^{j}\sum_{k = 1}^{m} kV_{i, j, k}^{m}.
\end{equation}

Putting the above together, we obtain
\begin{equation}
\label{Proof: Results for kV}
\left(\sum_{\alpha = 1}^{i}\frac{2^{\alpha - 1}x}{1 + 2^{\alpha - 1}x}\right) \prod_{\beta = 1}^{m}\left(1 + 2^{\beta - 1}x\right) = \sum_{j = 0}^{m}  2^{\frac{1}{2}j\left(j-1\right)} \genfrac{[}{]}{0pt}{0}{m}{j} x^{j}\sum_{k = 1}^{m} kV_{i, j, k}^{m}.
\end{equation}

Now we consider the following identity deduced from (\ref{Basic Identity Prod = Sum}) by replacing $x$ with $x^{-1}$,
\begin{equation}
\label{Basic Identity Prod = Sum, a variant form}
\prod_{\alpha = 1}^{m}\left(x + 2^{\alpha - 1}\right) = \sum_{j = 0}^{m}2^{\frac{1}{2}j\left(j-1\right)}\genfrac{[}{]}{0pt}{0}{m}{m - j}x^{m - j}, \quad \forall x \in \mathbb{R}.
\end{equation}

Similarly, for $x, y > 0$, define $g\left(x, y\right)$ as follows\begin{equation}
\label{Definition: g(x,y)}
g\left(x, y\right) = \prod_{\alpha = 1}^{i}\left(x + 2^{\alpha - 1}\right) \prod_{\beta = 1}^{m - i}\left(y + 2^{\beta + i - 1}\right).
\end{equation}

Again, applying (\ref{Basic Identity Prod = Sum, a variant form}) to (\ref{Definition: g(x,y)}) yields
\begin{align}
g\left(x, y\right) =& 2^{i\left(m - i\right)} \prod_{\alpha = 1}^{i}\left(x + 2^{\alpha - 1}\right) \prod_{\beta = 1}^{m - i}\left(2^{-i}y + 2^{\beta - 1}\right) \\
=& \sum_{l_{2} = 0}^{m - i}\sum_{l_{1} = 0}^{i} 2^{i\left(m - i\right) + \frac{1}{2}l_{1}\left(l_{1}-1\right) + \frac{1}{2}l_{2}\left(l_{2}-1\right) - i\left(m - i -l_{2}\right)} \genfrac{[}{]}{0pt}{0}{i}{i - l_{1}} \genfrac{[}{]}{0pt}{0}{m - i}{m - i - l_{2}} x^{i - l_{1}} y^{m - i - l_{2}} \\
\stackrel{(a)}{=}& \sum_{k = i}^{m} \sum_{j = k - i}^{k} 2^{\frac{1}{2}j\left(j-1\right) + \left(k-i\right)\left(k-j\right)}    \genfrac{[}{]}{0pt}{0}{i}{k - j} \genfrac{[}{]}{0pt}{0}{m - i}{m - k} x^{k - j} y^{m - k} \\
=& \sum_{j = 0}^{m}  2^{\frac{1}{2}j\left(j-1\right)} \genfrac{[}{]}{0pt}{0}{m}{m - j} \sum_{k = 0}^{m} C_{i, j, k}^{m} x^{k - j} y^{m - k}
\end{align}
where (a) is obtained by replacing $l_{1}$ and $l_{2}$ with $i + j - k$ and $k - i$, respectively.

Following the same procedure as in (\ref{Proof: Procedure to f(x,y) 1})-(\ref{Proof: Results for kV}), we can show that
\begin{align}
\left.\left[x \frac{\partial}{\partial x}g\left(x, y\right)\right]\right|_{y = x} =& \left(\sum_{\alpha = 1}^{i} \frac{x}{x + 2^{\alpha - 1}}\right) \prod_{\beta = 1}^{m}\left(x + 2^{\beta - 1}\right) \\
\label{Proof: Intermediate Results for kC}
=& \sum_{j = 0}^{m}  2^{\frac{1}{2}j\left(j-1\right)} \genfrac{[}{]}{0pt}{0}{m}{m - j} x^{m - j}\sum_{k = 0}^{m} \left(k - j\right)C_{i, j, k}^{m}.
\end{align}

Now, by substituting $\genfrac{[}{]}{0pt}{1}{m}{m - j} = \genfrac{[}{]}{0pt}{1}{m}{j}$ and $\sum\nolimits_{k = 0}^{m} C_{i, j, k}^{m} = 1$ (see (\ref{Prp: Sum of V and C over k}) in Appendix \ref{APP: Some Useful Properties of Basic Operators}) into (\ref{Proof: Intermediate Results for kC}) and replacing $x$ with $x^{-1}$, we can deduce that
\begin{equation}
\label{Proof: Results for kC}
\left(\sum_{\alpha = 1}^{i} \frac{1}{1 + 2^{\alpha - 1}x}\right) \prod_{\beta = 1}^{m}\left(1 + 2^{\beta - 1}x\right) = \sum_{j = 0}^{m}  2^{\frac{1}{2}j\left(j-1\right)} \genfrac{[}{]}{0pt}{0}{m}{j} x^{j}\sum_{k = 1}^{m} kC_{i, j, k}^{m} - \sum_{j = 0}^{m}  2^{\frac{1}{2}j\left(j-1\right)} \genfrac{[}{]}{0pt}{0}{m}{j} j x^{j}.
\end{equation}

Next, by rearranging and combining (\ref{Proof: Results for kV}) and (\ref{Proof: Results for kC}) as follows and substituting (\ref{Basic Identity Prod = Sum}) to the term $\prod_{\beta = 1}^{m}\left(1 + 2^{\beta - 1}x\right)$, we have
\begin{align}
&\sum_{j = 0}^{m}  2^{\frac{1}{2}j\left(j-1\right)} \genfrac{[}{]}{0pt}{0}{m}{j} \left[\sum_{k = 1}^{m} k\left(V_{i, j, k}^{m} + C_{i, j, k}^{m}\right)\right] x^{j} \nonumber\\
=& \sum_{\alpha = 1}^{i}\left( \frac{2^{\alpha - 1}x}{1 + 2^{\alpha - 1}x} + \frac{1}{1 + 2^{\alpha - 1}x} \right) \prod_{\beta = 1}^{m}\left(1 + 2^{\beta - 1}x\right)+ \sum_{j = 0}^{m}  2^{\frac{1}{2}j\left(j-1\right)} \genfrac{[}{]}{0pt}{0}{m}{j} j x^{j} \\
=& i \prod_{\beta = 1}^{m}\left(1 + 2^{\beta - 1}x\right) + \sum_{j = 0}^{m}  2^{\frac{1}{2}j\left(j-1\right)} \genfrac{[}{]}{0pt}{0}{m}{j} j x^{j} = \sum_{j = 0}^{m}  2^{\frac{1}{2}j\left(j-1\right)} \genfrac{[}{]}{0pt}{0}{m}{j} \left(i + j\right) x^{j}.
\end{align}
Therefore
\begin{equation}
\label{Proof: kV + kC = i + j 1}
\sum_{k = 1}^{m} k\left(V_{i, j, k}^{m} + C_{i, j, k}^{m}\right) = i + j, \quad \forall i, j \in \mathbb{M}.
\end{equation}

Finally, by substituting (\ref{Proof: kV + kC = i + j 1}) to (\ref{Proof: kV + kC = i + j}){}, we complete the proof of Lemma \ref{LEM: Duality Rule for Entropy}.
\end{IEEEproof}

\begin{remark}
Now let us briefly discuss (\ref{Proof: kV + kC = i + j 1}) and interpret the operational meaning of the duality rule for entropy (\ref{Lemma: Duality Rule for Entropy}).
For any fixed $i, j \in \mathbb{M}$, consider two statistically independent messages, the a posteriori PMFs of which are equiprobable over a subspace $\mathcal{S}$ of dimension $i$ and a subspace $\mathcal{S}'$ of dimension $j$, respectively.
As discussed in Remark \ref{RMK: Discussion On H}, the uncertainties of the two messages are given by $i$ bits and $j$ bits, and therefore, the total uncertainty is given by $i + j$ bits.
If we combine the two messages based on the VN (resp. CN) decoding algorithm, then the a posteriori PMF of the combined message is equiprobable over the intersection (resp. sum) of $\mathcal{S}$ and $\mathcal{S}'$, denoted as $\mathcal{S} \cap \mathcal{S}'$ (resp. $\mathcal{S} + \mathcal{S}'$).
Moreover, as interpreted in Subsection II-A in \cite{7430313}, $V_{i, j, k}^{m}$ (resp. $C_{i, j, k}^{m}$) is the probability of the event that the dimension of $\mathcal{S} \cap \mathcal{S}'$ (resp. $\mathcal{S} + \mathcal{S}'$), or equivalently, the uncertainty of the combined message, is exactly $k$ (bits).
Therefore, the identity (\ref{Proof: kV + kC = i + j 1}) indicates that the total (average) uncertainty is invariant under the combinations of two statistically independent messages based on the VN and the CN decoding algorithms.
This explains why we mentioned in the introduction that the duality rule for entropy (\ref{Lemma: Duality Rule for Entropy}) reveals a conservation relation between the input and the output entropies of VNs and CNs.
\end{remark}

Following the same line as in \cite{6912949}, we extend the rule (\ref{Lemma: Duality Rule for Entropy}) to the following relations and omit the details for brevity.
\begin{corollary}
\label{COL: Duality Rule for Entropy Ext}
For any $\mathtt{a}, \mathtt{b}, \mathtt{c}, \mathtt{d} \in \mathcal{X}$,
\begin{equation}
\label{Corollary: Duality Rule for Entropy Ext1}
H\left(\mathtt{a}\boxdot\left(\mathtt{b} - \mathtt{c}\right)\right) + H\left(\mathtt{a}\boxtimes\left(\mathtt{b} - \mathtt{c}\right)\right) = H\left(\mathtt{b} - \mathtt{c}\right),
\end{equation}
\begin{equation}
\label{Corollary: Duality Rule for Entropy Ext2}
H\left(\left(\mathtt{a}-\mathtt{b}\right)\boxdot\left(\mathtt{c} - \mathtt{d}\right)\right) + H\left(\left(\mathtt{a}-\mathtt{b}\right)\boxtimes\left(\mathtt{c} - \mathtt{d}\right)\right) = 0.
\end{equation}
\end{corollary}

\subsection{Partial Ordering}
\label{Partial Ordering}
An important issue in the DE analysis is comparing two densities to identify which one offers more information about the coded symbols.
For this purpose, a concept termed partial ordering is established in \cite{Richardson:2008:MCT:1795974} based on statistical degradation in the context of binary LDPC ensembles over BMS channels.
In \cite{7430313}, the authors defined partial ordering based on the complementary cumulative distribution function to accommodate the analysis of nonbinary LDPC ensembles on the BEC.
We will exploit the notion of partial ordering in \cite{7430313} in this paper, the definition of which is reformulated as follows.

\begin{definition}
\label{DEF: Partial Order}
For any $\mathtt{a}, \mathtt{b} \in \mathcal{X}$, we say that $\mathtt{a} \preceq \mathtt{b}$ or $\mathtt{b} \succeq \mathtt{a}$ if the following inequality holds
\begin{equation}
\label{Definition: Partial Order}
\sum_{n = k}^{m} a_{n} \le \sum_{n = k}^{m} b_{n}, \quad \forall k \in \mathbb{M}\backslash\left\{0\right\}
\end{equation}
and say that $\mathtt{a} \prec \mathtt{b}$ or $\mathtt{b} \succ \mathtt{a}$ if $\mathtt{a} \preceq \mathtt{b}$ and $\mathtt{a} \neq \mathtt{b}$.
\end{definition}

\begin{proposition}
\label{PRP: another Def of strict partial ordering}
For any $\mathtt{a}, \mathtt{b} \in \mathcal{X}$, the strict partial order $\mathtt{a} \prec \mathtt{b}$ holds if and only if there exists a nonempty set $\mathcal{I} \subseteq \mathbb{M}\backslash\left\{0\right\}$ such that $\sum_{n = k}^{m} a_{n} < \sum_{n = k}^{m} b_{n}$ for $k \in \mathcal{I}$ and $\sum_{n = k}^{m} a_{n} = \sum_{n = k}^{m} b_{n}$ for $k \notin \mathcal{I}$.
\end{proposition}

\begin{IEEEproof}
The proof is straightforward and we omit it for simplicity.
\end{IEEEproof}

It is easy to justify that $\mathtt{\Delta}_{0} \preceq \mathtt{a} \preceq \mathtt{\Delta}_{m}, \forall \mathtt{a} \in \mathcal{X}$.

\begin{proposition}
\label{PRP: Convergence of Partially Ordered Densitiy}
Consider a series of densities $\left\{\mathtt{x}_{l}\right\}_{l \in \mathbb{Z}}$.
The limit $\lim_{l \to \infty}\mathtt{x}_{l}$ exists if either $\mathtt{x}_{l + 1} \preceq \mathtt{x}_{l}$ or $\mathtt{x}_{l + 1} \succeq \mathtt{x}_{l}$ holds for all $l \in \mathbb{Z}$.
\end{proposition}

\begin{IEEEproof}
By definition, we can deduce that either $\sum_{n = k}^{m}\left[\mathtt{x}_{l + 1}\right]_{n} \le \sum_{n = k}^{m}\left[\mathtt{x}_{l}\right]_{n}$ or $\sum_{n = k}^{m}\left[\mathtt{x}_{l + 1}\right]_{n} \ge \sum_{n = k}^{m}\left[\mathtt{x}_{l}\right]_{n}$ holds for each $k \in \mathbb{M}\backslash\left\{0\right\}$.
Notice that $\sum_{n = k}^{m}\left[\mathtt{x}_{l}\right]_{n}$ is always bounded between $0$ and $1$.
Therefore, the limit $\lim_{l \to \infty}\mathtt{x}_{l}$ does indeed exist.
\end{IEEEproof}

\begin{proposition}
\label{PRP: PotFun Preserves the Partial Order}
The entropy function $H\left(\cdot\right)$ preserves partial ordering. More precisely, for any $\mathtt{a}, \mathtt{b} \in \mathcal{X}$, we have $H\left(\mathtt{a}\right) \le H\left(\mathtt{b}\right)$ if $\mathtt{a} \preceq \mathtt{b}$, and $H\left(\mathtt{a}\right) < H\left(\mathtt{b}\right)$ if $\mathtt{a} \prec \mathtt{b}$.
\end{proposition}

\begin{IEEEproof}
By Definition \ref{DEF: Partial Order}, $\mathtt{a} \preceq \mathtt{b}$ implies that $\sum_{n = k}^{m} a_{n} \le \sum_{n = k}^{m} b_{n}, \forall k \in \mathbb{M} \backslash \left\{0\right\}$. Therefore
\begin{equation}
H\left(\mathtt{a}\right) = \sum_{k = 0}^{m} k a_{k} \stackrel{(a)}{=} \sum_{k = 1}^{m} \sum_{n = k}^{m} a_{n} \le \sum_{k = 1}^{m} \sum_{n = k}^{m} b_{n} = \sum_{k = 0}^{m} k b_{k} = H\left(\mathtt{b}\right)
\end{equation}
where (a) is based on Proposition \ref{PRP: An important TRICK} in Appendix \ref{APP: A Useful Tool}.

The proof of the implication $\mathtt{a} \prec \mathtt{b} \Rightarrow H\left(\mathtt{a}\right)  < H\left(\mathtt{b}\right)$ now becomes straightforward by Proposition \ref{PRP: another Def of strict partial ordering}.
\end{IEEEproof}


\begin{lemma}
\label{LEM: Basic Operators Preserve the Partial Order}
The VN and CN convolutional operators $\boxdot$ and $\boxtimes$ preserve partial ordering. More precisely, for any $\mathtt{a}, \mathtt{b}, \mathtt{c} \in \mathcal{X}$ with $\mathtt{a} \preceq \mathtt{b}$, we have
\begin{equation}
\label{Lemma: Basic Operators Preserve the Partial Order with =}
\mathtt{a} \boxdot \mathtt{c} \preceq \mathtt{b} \boxdot \mathtt{c} \quad \text{and} \quad \mathtt{a} \boxtimes \mathtt{c} \preceq \mathtt{b} \boxtimes \mathtt{c}.
\end{equation}
Further, if $\mathtt{a} \prec \mathtt{b}$, then
\begin{equation}
\label{Lemma: Basic Operators Preserve the Partial Order 1}
\mathtt{a} \boxdot \mathtt{c} \prec \mathtt{b} \boxdot \mathtt{c}, \quad \text{for } \mathtt{c} \neq \mathtt{\Delta}_{0},\text{ }
\end{equation}
\begin{equation}
\label{Lemma: Basic Operators Preserve the Partial Order 2}
\mathtt{a} \boxtimes \mathtt{c} \prec \mathtt{b} \boxtimes \mathtt{c}, \quad \text{for } \mathtt{c} \neq \mathtt{\Delta}_{m}.
\end{equation}
\end{lemma}
\begin{IEEEproof}
We focus on the results for $\boxdot$.
The proof for $\boxtimes$ is identical by noticing that $C_{i, j, k}^{m} = V_{m - i, m - j, m - k}^{m}$.

Notice that, for $n \in \mathbb{M}\backslash\left\{0\right\}$, $\mathtt{a} \preceq \mathtt{b}$ implies $\sum_{k = n} a_{k} \le \sum_{k = n} b_{k}$ by Definition \ref{DEF: Partial Order}. Therefore
\begin{align}
\label{Proof: Basic Operators Preserve the Partial Order}
\sum_{k = n}^{m} \left[\mathtt{a} \boxdot \mathtt{c}\right]_{k} =& \sum_{i = 0}^{m} \sum_{j = 0}^{m} \sum_{k = n}^{m} a_{i} V_{i, j, k}^{m} c_{j} \stackrel{(a)}{=} \sum_{j = 0}^{m} c_{j} \sum_{k = n}^{m}  \sum_{i = n}^{m} a_{i} V_{i, j, k}^{m} \\
\stackrel{(b)}{=}& \sum_{j = 0}^{m} c_{j} \sum_{k = n}^{m} \left[ V_{n, j, k}^{m} \sum_{l = n}^{m} a_{l} + \sum_{i = n + 1}^{m}  \left(V_{i, j, k}^{m} - V_{i-1, j, k}^{m}\right)\sum_{l = i}^{m} a_{l} \right] \\
\stackrel{(c)}{=}& \sum_{j = 0}^{m} c_{j} \sum_{i = n}^{m} \left[ \sum_{k = n}^{m} \left(V_{i, j, k}^{m} - V_{i-1, j, k}^{m}\right)  \right] \sum_{l = i}^{m} a_{l} \\
\stackrel{(d)}{\le}& \sum_{j = 0}^{m} c_{j} \sum_{i = n}^{m} \left[ \sum_{k = n}^{m} \left(V_{i, j, k}^{m} - V_{i-1, j, k}^{m}\right)  \right] \sum_{l = i}^{m} b_{l} = \sum_{k = n}^{m} \left[\mathtt{b} \boxdot \mathtt{c}\right]_{k}
\end{align}
where (a) and (c) are based on the fact that $V_{i, j, k}^{m} = 0$ if $0 \le i < n \le k$, (b) follows from Proposition \ref{PRP: An important TRICK} in Appendix \ref{APP: A Useful Tool}, and (d) is based on Claim 3) of Proposition \ref{PRP: Properties of V and C} in Appendix \ref{APP: Some Useful Properties of Basic Operators}.
Therefore, by Definition \ref{DEF: Partial Order}, we obtain the desired result $\mathtt{a} \boxdot \mathtt{c} \preceq \mathtt{b} \boxdot \mathtt{c}$.

Next, we show that $\boxdot$ preserves strict partial ordering. To this end, we rearrange the above as follows
\begin{equation}
\label{Proof: Basic Operators Preserve the Partial Order 2}
\sum_{k = n}^{m} \left[\mathtt{a} \boxdot \mathtt{c}\right]_{k} - \sum_{k = n}^{m} \left[\mathtt{b} \boxdot \mathtt{c}\right]_{k} = \sum_{j = 0}^{m} c_{j} \sum_{i = n}^{m} \left[ \sum_{k = n}^{m}  \left(V_{i, j, k}^{m} - V_{i-1, j, k}^{m} \right) \right] \left(\sum_{l = i}^{m} a_{l} - \sum_{l = i}^{m} b_{l}\right).
\end{equation}

Let $i_0, j_0 \in \mathbb{M}\backslash\left\{0\right\}$ be two integers satisfying $\sum_{l = i_0}^{m} a_{l} - \sum_{l = i_0}^{m} b_{l} < 0$ and $c_{j_0} > 0$.
The existence of such $i_0$ and $j_0$ is guaranteed by the assumption $\mathtt{a} \prec \mathtt{b}$ and $\mathtt{c} \neq \mathtt{\Delta}_{0}$.
By letting $n = \min\left\{i_0, j_0\right\}$ and discarding some nonpositive terms in (\ref{Proof: Basic Operators Preserve the Partial Order 2}), we can obtain a strictly negative upper bound, i.e.,
\begin{equation}
\sum_{k = n}^{m} \left[\mathtt{a} \boxdot \mathtt{c}\right]_{k} - \sum_{k = n}^{m} \left[\mathtt{b} \boxdot \mathtt{c}\right]_{k} \le c_{j_0} \left[ \sum_{k = n}^{m}  \left(V_{i_0, j_0, k}^{m} - V_{i_0-1, j_0, k}^{m}\right) \right] \left(\sum_{l = i_0}^{m} a_{l} - \sum_{l = i_0}^{m} b_{l}\right) < 0.
\end{equation}

Thus $\sum_{k = n}^{m} \left[\mathtt{a} \boxdot \mathtt{c}\right]_{k} < \sum_{k = n}^{m} \left[\mathtt{b} \boxdot \mathtt{c}\right]_{k}$ holds for at least one integer $n \in \mathbb{M}\backslash\left\{0\right\}$.
This completes the proof of (\ref{Lemma: Basic Operators Preserve the Partial Order 1}).
\end{IEEEproof}

\begin{remark}
The partial order preservation property in Lemma \ref{LEM: Basic Operators Preserve the Partial Order} guarantees that, if an uncoupled DE recursion (see (\ref{Definition: Uncoupled DE 2}) in the sequel) is initialized by $\mathtt{\Delta}_{m}$, i.e., the most ``uncertain'' density, then the densities generated by the DE recursion are always partially ordered as the iteration proceeds.
\end{remark}

\begin{lemma}
\label{LEM: Conv of Diff Densities}
For any $\mathtt{a}, \mathtt{b}, \mathtt{c}, \mathtt{d} \in \mathcal{X}$ with $\mathtt{a} \preceq \mathtt{b}$ and $\mathtt{c} \preceq \mathtt{d}$, we have
\begin{equation}
\label{Lemma: Conv of Diff Densities}
H\left(\left(\mathtt{a} - \mathtt{b}\right)\boxdot \left(\mathtt{c} - \mathtt{d} \right)\right) \ge  0, \quad
H\left(\left(\mathtt{a} - \mathtt{b}\right) \boxtimes \left(\mathtt{c} - \mathtt{d}\right)\right) \le 0
\end{equation}
where the equalities hold if and only if $\mathtt{a} = \mathtt{b}$ and $\mathtt{c} = \mathtt{d}$.
\end{lemma}
\begin{IEEEproof}
For any $i, n \in \mathbb{M}\backslash\left\{0\right\}, j \in \mathbb{M}$, we define
\begin{equation}
D_{i, j, n}^{m} = \frac{2^{\left(i - n\right)\left(j - n + 1\right)}\genfrac{[}{]}{0pt}{0}{i - 1}{n - 1}\genfrac{[}{]}{0pt}{0}{m - i}{j - n}}{\genfrac{[}{]}{0pt}{0}{m}{j}}.
\end{equation}

Obviously, $D_{i, j, n}^{m} > 0$ if $n < i$ and $i + j \le m + n$.
Further, if $j > 0$, then
\begin{equation}
\label{Proof: Dijn is increasing w.r.t j}
D_{i, j, n}^{m} = D_{i, j-1, n}^{m} \frac{\left(2^{j - n} - 2^{- n}\right)\left(2^{m - j + n + 1} - 2^{i}\right)}{\left(2^{j - n} - 1\right)\left(2^{m - j + 1} - 1\right)}  > D_{i, j-1, n}^{m} 2^{n - 1} > D_{i, j-1, n}^{m}.
\end{equation}

On the other hand, based on (\ref{Proof: Results for kV}), for $i \in \mathbb{M}\backslash\left\{0\right\}$, we have
\begin{align}
&\sum_{j = 0}^{m}  2^{\frac{1}{2}j\left(j-1\right)} \genfrac{[}{]}{0pt}{0}{m}{j} x^{j}\sum_{k = 1}^{m} k \left(V_{i, j, k}^{m} - V_{i - 1, j, k}^{m}\right) \nonumber \\
=& \frac{2^{i - 1}x}{1 + 2^{i - 1}x} \prod_{\beta = 1}^{m}\left(1 + 2^{\beta - 1}x\right) = 2^{i - 1}x \prod_{\alpha = 1}^{i - 1}\left(1 + 2^{\alpha - 1}x\right) \prod_{\beta = 1}^{m - i}\left(1 + 2^{\beta - 1} \left(2^{i}x\right)\right) \\
\stackrel{(a)}{=}& \sum_{l_{1} = 0}^{i - 1}\sum_{l_{2} = 0}^{i - 1} 2^{i - 1 + \frac{1}{2}l_{1}\left(l_{1} - 1\right) + \frac{1}{2}l_{1}\left(l_{1} - 1\right) + il_{2}}  \genfrac{[}{]}{0pt}{0}{i - 1}{l_{1}} \genfrac{[}{]}{0pt}{0}{m - i}{l_{2}} x^{l_{1} + l_{2} + 1} \\
\stackrel{(b)}{=}& \sum_{j = 0}^{m} 2^{\frac{1}{2}j\left(j - 1\right)} x^{j} \sum_{n = 1}^{m} 2^{\left(i  - n\right)\left(j - n + 1\right)} \genfrac{[}{]}{0pt}{0}{i - 1}{n - 1} \genfrac{[}{]}{0pt}{0}{m - i}{j - n} = \sum_{j = 0}^{m} 2^{\frac{1}{2}j\left(j - 1\right)} \genfrac{[}{]}{0pt}{0}{m}{j} x^{j} \sum_{n = 1}^{m} D_{i, j, n}^{m}
\end{align}
where (a) is based on (\ref{Basic Identity Prod = Sum}) and (b) is obtained by $l_{1} = n - 1$ and $l_{2} = j - n$.
Therefore, we have
\begin{equation}
\label{Proof: Gap of kVijk wrt i}
\sum_{k = 1}^{m} k \left(V_{i, j, k}^{m} - V_{i - 1, j, k}^{m}\right)  = \sum_{n = 1}^{m} D_{i, j, n}^{m} \ge 0.
\end{equation}

Following (\ref{Proof: Dijn is increasing w.r.t j}), we have
\begin{equation}
\label{Proof: PD of Oder2 for SumkV}
\sum_{k = 1}^{m}k\left(V_{i, j, k}^{m} - V_{i - 1, j, k}^{m} - V_{i, j - 1, k}^{m} + V_{i - 1, j - 1, k}^{m} \right) = \sum_{n = 1}^{m} \left(D_{i, j, n}^{m} - D_{i, j - 1, n}^{m}\right) > 0.
\end{equation}

By Definition \ref{DEF: Partial Order}, $\forall i, j \in \mathbb{M}\backslash\left\{0\right\}$, $\mathtt{a} \preceq \mathtt{b}$ and $\mathtt{c} \preceq \mathtt{d}$ imply $\sum_{n = i}^{m} \left(a_{n} - b_{n}\right) \le 0$ and $\sum_{l = j}^{m} \left(c_{l} - d_{l}\right) \le 0$, respectively.
The first inequality in (\ref{Lemma: Conv of Diff Densities}) can be deduced from (\ref{Proof: PD of Oder2 for SumkV}).
Specifically,
\begin{align}
&H\left(\left(\mathtt{a} - \mathtt{b}\right)\boxdot \left(\mathtt{c} - \mathtt{d} \right)\right) \nonumber \\
=& \sum_{k = 1}^{m}k \sum_{j = k}^{m} \left(c_{j} - d_{j}\right) \sum_{i = k}^{m} \left(a_{i} - b_{i}\right) V_{i, j, l}^{m}\\
=& \sum_{k = 1}^{m}k \sum_{j = k}^{m} \left(c_{j} - d_{j}\right) \sum_{i = k}^{m} \left(V_{i, j, k}^{m} - V_{i - 1, j, k}^{m}\right) \sum_{n = i}^{m} \left(a_{n} - b_{n}\right) \\
=& \sum_{k = 1}^{m}k \sum_{i = k}^{m} \sum_{j = k}^{m} \left(V_{i, j, k}^{m} - V_{i - 1, j, k}^{m} - V_{i, j - 1, k}^{m} + V_{i - 1, j - 1, k}^{m}\right) \sum_{n = i}^{m} \left(a_{n} - b_{n}\right) \sum_{l = j}^{m} \left(c_{l} - d_{l}\right) \\
\label{Proof: Conv of Diff Densities 1}
=& \sum_{i = 1}^{m} \sum_{j = 1}^{m} \left[\sum_{n = i}^{m} \left(a_{n} - b_{n}\right)\right] \left[\sum_{l = j}^{m} \left(c_{l} - d_{l}\right)\right] \sum_{k = 1}^{m}k\left(V_{i, j, k}^{m} - V_{i - 1, j, k}^{m} - V_{i, j - 1, k}^{m} + V_{i - 1, j - 1, k}^{m}\right) \ge 0.
\end{align}

Further, due to the strict positiveness of the right-hand side of (\ref{Proof: PD of Oder2 for SumkV}), the equality in (\ref{Proof: Conv of Diff Densities 1}) holds if and only if $\mathtt{a} = \mathtt{b}$ and $\mathtt{c} = \mathtt{d}$.

Now, based on (\ref{Corollary: Duality Rule for Entropy Ext2}), the proof of the second inequality in (\ref{Lemma: Conv of Diff Densities}) becomes straightforward.
\end{IEEEproof}


The following corollary simply follows from Lemma \ref{LEM: Conv of Diff Densities}.
\begin{corollary}
\label{COL: Conv of Square Diff Densities}
For any $\mathtt{a}, \mathtt{b}, \mathtt{c} \in \mathcal{X}$, we have
\begin{equation}
\label{Lemma: Conv of Diff Densities 3}
H\left(\left(\mathtt{a} - \mathtt{b}\right)^{\boxdot 2} \boxdot \mathtt{c}\right) \ge 0, \quad
H\left(\left(\mathtt{a} - \mathtt{b}\right)^{\boxtimes 2} \boxtimes \mathtt{c}\right) \le 0
\end{equation}
if either $\mathtt{a} \preceq \mathtt{b}$ or $\mathtt{a} \succeq \mathtt{b}$ holds.
Moreover, the equalities in the above hold if and only if $\mathtt{a} = \mathtt{b}$.
\end{corollary}

\begin{lemma}
\label{LEM: UB for Conv of Diff Densities}
For any $\mathtt{a}, \mathtt{b}, \mathtt{c}, \mathtt{d} \in \mathcal{X}$ with $\mathtt{a} \succeq \mathtt{b}$, we have
\begin{align}
\left|H\left(\left(\mathtt{a} - \mathtt{b}\right) * \left(\mathtt{c} - \mathtt{d} \right)\right)\right| \le H\left(\mathtt{a} - \mathtt{b}\right).
\end{align}
\end{lemma}
\begin{IEEEproof}
We rewrite (\ref{Proof: Conv of Diff Densities 1}) as follows.
\begin{align}
\label{Proof: UB for Conv of Diff Densities}
H\left(\left(\mathtt{a} - \mathtt{b}\right)\boxdot \left(\mathtt{c} - \mathtt{d} \right)\right) =& \sum_{j = 0}^{m} c_{j} \sum_{i = 0}^{m} \sum_{k = 1}^{m}k\left(V_{i, j, k}^{m} - V_{i - 1, j, k}^{m}\right) \sum_{n = i}^{m} \left(a_{n} - b_{n}\right) \nonumber \\
-& \sum_{j = 0}^{m} d_{j} \sum_{i = 0}^{m} \sum_{k = 1}^{m}k\left(V_{i, j, k}^{m} - V_{i - 1, j, k}^{m}\right) \sum_{n = i}^{m} \left(a_{n} - b_{n}\right).
\end{align}

By assumption $\mathtt{a} \succeq \mathtt{b}$ and (\ref{Proof: Gap of kVijk wrt i}), it is obvious that the four terms $\sum_{n = i}^{m} \left(a_{n} - b_{n}\right)$, $\sum_{k = 1}^{m}k(V_{i, j, k}^{m} - V_{i - 1, j, k}^{m})$, $c_{j}$ and $d_{j}$ are always nonnegative, satisfying $\sum_{j = 0}^{m} c_{j} = \sum_{j = 0}^{m} d_{j} = 1$.
Further, the inequality (\ref{Proof: PD of Oder2 for SumkV}) indicates that the second term $\sum_{k = 1}^{m} k (V_{i, j, k}^{m} - V_{i - 1, j, k}^{m})$ is strictly increasing with respect to $j \in \mathbb{M}$.
Therefore, $H\left(\left(\mathtt{a} - \mathtt{b}\right) \boxdot \left(\mathtt{c} - \mathtt{d} \right)\right)$ is maximized by $c_{m} = d_{0} = 1$ (i.e., $\mathtt{c} = \mathtt{\Delta}_{m}, \mathtt{d} = \mathtt{\Delta}_{0}$) and minimized by $c_{0} = d_{m} = 1$ (i.e., $\mathtt{c} = \mathtt{\Delta}_{0}, \mathtt{d} = \mathtt{\Delta}_{m}$).
This leads to the following desired result, i.e.,
\begin{equation}
- H\left(\mathtt{a} - \mathtt{b}\right) \le - H\left(\left(\mathtt{a} - \mathtt{b}\right)\boxtimes \left(\mathtt{c} - \mathtt{d} \right)\right) = H\left(\left(\mathtt{a} - \mathtt{b}\right)\boxdot \left(\mathtt{c} - \mathtt{d} \right)\right) \le H\left(\mathtt{a} - \mathtt{b}\right).
\end{equation}
\end{IEEEproof}

\section{Potential Functions and Threshold Saturation}
\label{Potential Analysis and Threshold Saturation}

\subsection{LDPC$\left(\lambda, \rho, {m}\right)$}
\begin{definition}
\label{DEF: Uncoupled DE}
For LDPC$\left(\lambda, \rho, {m}\right)$ on the BEC with erasure probability $\epsilon \in  \left[0, 1\right]$, the uncoupled DE recursion in the $l$-th iteration is given by
\begin{equation}
\label{Definition: Uncoupled DE 1}
\begin{cases}
\mathtt{y}^{\left(l\right)} = \rho^{\boxtimes} \left(\mathtt{x}^{\left(l - 1\right)}\right) \\
\mathtt{x}^{\left(l\right)} = \mathtt{c}_{\epsilon} \boxdot \lambda^{\boxdot} \left( \mathtt{y}^{\left(l\right)}\right)
\end{cases}
\forall l \in \mathbb{Z}_{+}
\end{equation}
or equivalently,
\begin{equation}
\label{Definition: Uncoupled DE 2}
\mathtt{x}^{\left(l\right)} = \mathtt{c}_{\epsilon} \boxdot \lambda^{\boxdot} \left( \rho^{\boxtimes} \left(\mathtt{x}^{\left(l - 1\right)}\right) \right), \quad \forall l \in \mathbb{Z}_{+}
\end{equation}
where $\mathtt{x}^{\left(l\right)}$ and $\mathtt{y}^{\left(l\right)}$ are respectively the output densities of VNs and CNs, $\mathtt{c}_{\epsilon}$ the channel density,
and the operators $\lambda^{\boxdot}\left(\cdot\right)$ and $\rho^{\boxtimes}\left(\cdot\right)$ defined as
\begin{equation}
\label{Definition: Transfer Operators at VNs and CNs}
\lambda^{\boxdot}\left(\mathtt{a}\right) = \sum_{i} \lambda_i \mathtt{a}^{\boxdot {i - 1}}, \quad \rho^{\boxtimes}\left(\mathtt{b}\right) = \sum_{j} \rho_j \mathtt{b}^{\boxtimes {j - 1}}
\end{equation}
$\forall \mathtt{a}, \mathtt{b} \in \mathcal{X}$.
Here, the $k$-th entry of the channel density $\mathtt{c}_{\epsilon}$ is determined by
\begin{equation}
\label{Definition: Channel Density}
\left[\mathtt{c}_{\epsilon}\right]_{k} = \genfrac{(}{)}{0pt}{0}{m}{k} \epsilon^{k}\left(1- \epsilon\right)^{m - k},    \quad k \in \mathbb{M}.
\end{equation}
\end{definition}

As in \cite{6912949}, $\forall l \in \mathbb{Z}$, we will write $\mathtt{x}^{\left(l\right)}$ in the form of $\mathtt{x}^{\left(l\right)} = \mathtt{T}_\text{s}^{\left(l\right)}\left(\mathtt{x}^{\left(0\right)}, \mathtt{c}_{\epsilon}\right)$ to emphasize that the output density at VNs in the $l$-th iteration is determined by the initial density $\mathtt{x}^{\left(0\right)}$ and the channel density $\mathtt{c}_{\epsilon}$.
Moreover, based on those propositions and lemmas in Subsection \ref{Partial Ordering}, we can show that the DE operator $\mathtt{T}_\text{s}^{\left(l\right)}\left(\cdot, \cdot\right)$ satisfies the same monotonicity properties as stated in Lemma 18 in \cite{6912949}.
For convenience, we reformulate these properties in the following lemma.

\begin{lemma}
\label{LEM: Monotonicity of Ts}
For any $l \in \mathbb{Z}_{+}$ and $\mathtt{a}_1, \mathtt{a}_2, \mathtt{a}, \mathtt{c}_1, \mathtt{c}_2, \mathtt{c} \in \mathcal{X}$, the DE update operator $\mathtt{T}_\text{s}^{\left(l\right)}\left(\cdot, \cdot\right)$ satisfies the following properties.

1) $\mathtt{T}_\text{s}^{\left(l\right)}\left(\mathtt{a}_1, \mathtt{c}\right) \succeq \mathtt{T}_\text{s}^{\left(l\right)}\left(\mathtt{a}_2, \mathtt{c}\right)$ if $\mathtt{a}_1 \succeq \mathtt{a}_2$.

2) $\mathtt{T}_\text{s}^{\left(l\right)}\left(\mathtt{a}, \mathtt{c}_1\right) \succeq \mathtt{T}_\text{s}^{\left(l\right)}\left(\mathtt{a}, \mathtt{c}_2\right)$ if $\mathtt{c}_1 \succeq \mathtt{c}_2$.

3) If $\exists \mathtt{a} \in \mathcal{X}$ such that $\mathtt{T}_\text{s}^{\left(1\right)}\left(\mathtt{a}, \mathtt{c}\right) \preceq \mathtt{a}$, then $\mathtt{T}_\text{s}^{\left(l + 1\right)}\left(\mathtt{a}, \mathtt{c}\right) \preceq \mathtt{T}_\text{s}^{\left(l\right)}\left(\mathtt{a}, \mathtt{c}\right)$ and the limit $\mathtt{T}_\text{s}^{\left(\infty\right)}\left(\mathtt{a}, \mathtt{c}\right) = \lim_{l \to \infty} \mathtt{T}_\text{s}^{\left(l\right)}\left(\mathtt{a}, \mathtt{c}\right)$ does indeed exist, satisfying $\mathtt{T}_\text{s}^{\left(\infty\right)}\left(\mathtt{a}, \mathtt{c}\right) \preceq \mathtt{T}_\text{s}^{\left(l\right)}\left(\mathtt{a}, \mathtt{c}\right)$ and
\begin{equation}
\mathtt{T}_\text{s}^{\left(1\right)}\left(\mathtt{T}_\text{s}^{\left(\infty\right)}\left(\mathtt{a}, \mathtt{c}\right), \mathtt{c}\right )= \mathtt{T}_\text{s}^{\left(\infty\right)}\left(\mathtt{a}, \mathtt{c}\right).
\end{equation}

4) If $\exists \mathtt{a} \in \mathcal{X}$ such that $\mathtt{T}_\text{s}^{\left(1\right)}\left(\mathtt{a}, \mathtt{c}\right) \succeq \mathtt{a}$, then $\mathtt{T}_\text{s}^{\left(l + 1\right)}\left(\mathtt{a}, \mathtt{c}\right) \succeq \mathtt{T}_\text{s}^{\left(l\right)}\left(\mathtt{a}, \mathtt{c}\right)$ and the limit $\mathtt{T}_\text{s}^{\left(\infty\right)}\left(\mathtt{a}, \mathtt{c}\right) = \lim_{l \to \infty} \mathtt{T}_\text{s}^{\left(l\right)}\left(\mathtt{a}, \mathtt{c}\right)$ does indeed exist, satisfying $\mathtt{T}_\text{s}^{\left(\infty\right)}\left(\mathtt{a}, \mathtt{c}\right) \succeq \mathtt{T}_\text{s}^{\left(l\right)}\left(\mathtt{a}, \mathtt{c}\right)$ and
\begin{equation}
\mathtt{T}_\text{s}^{\left(1\right)}\left(\mathtt{T}_\text{s}^{\left(\infty\right)}\left(\mathtt{a}, \mathtt{c}\right), \mathtt{c}\right) = \mathtt{T}_\text{s}^{\left(\infty\right)}\left(\mathtt{a}, \mathtt{c}\right).
\end{equation}
\end{lemma}

\begin{IEEEproof}
The proof is identical to that of Lemma 18 in \cite{6912949} and we omit the details here.
\end{IEEEproof}

\begin{definition}
\label{DEF: Underlying Fixed Point}
For a fixed $\epsilon \in \left[0, 1\right]$, a density $\mathtt{x} \in \mathcal{X}$ is said to be an uncoupled fixed point (UFP) of the uncoupled DE recursion (\ref{Definition: Uncoupled DE 2}) if it satisfies $\mathtt{x} = \mathtt{T}_\text{s}^{\left(1\right)}\left(\mathtt{x}, \mathtt{c}_{\epsilon}\right)$.
In the sequel, we will use $\mathcal{F}_{\text{s}}\left(\epsilon\right)$ to denote the set of all such UFPs.
\end{definition}


\begin{definition}
\label{DEF: PotFun for UDE}
For any $\mathtt{x} \in \mathcal{X}$ and $\epsilon \in \left[0, 1\right]$, the potential function for LDPC$\left(\lambda, \rho, {m}\right)$ is given by
\begin{equation}
\label{Definition: PotFun for UDE}
U_{\text{s}}\left(\mathtt{x}, \epsilon\right) = \frac{L'\left(1\right)}{R'\left(1\right)}H\left(R^{\boxtimes}\left(\mathtt{x}\right)\right) + L'\left(1\right)H\left(\rho^{\boxtimes}\left(\mathtt{x}\right)\right) - L'\left(1\right)H\left(\mathtt{x} \boxtimes \rho^{\boxtimes}\left(\mathtt{x}\right)\right) - H\left(\mathtt{c}_{\epsilon}\boxdot L^{\boxdot}\left(\rho^{\boxtimes}\left(\mathtt{x}\right)\right)\right).
\end{equation}
\end{definition}

\begin{lemma}
\label{LEM: Monotonicity of Us}
If $\exists \mathtt{a} \in \mathcal{X}$ such that either $\mathtt{T}_\text{s}^{\left(1\right)}\left(\mathtt{a}, \mathtt{c}\right) \preceq \mathtt{a}$ or $\mathtt{T}_\text{s}^{\left(1\right)}\left(\mathtt{a}, \mathtt{c}\right) \succeq \mathtt{a}$ holds, then
\begin{equation}
\label{Lemma: Monotonicity of Us}
U_{\text{s}}\left(\mathtt{T}_\text{s}^{\left(l + 1\right)}\left(\mathtt{a}, \mathtt{c}\right), \epsilon\right) \le
U_{\text{s}}\left(\mathtt{T}_\text{s}^{\left(l\right)}\left(\mathtt{a}, \mathtt{c}\right), \epsilon\right), \forall l \in \mathbb{Z}.
\end{equation}
\end{lemma}
\begin{IEEEproof}
For notational brevity, we define
\begin{equation}
\label{Definition: Ws 1}
W_{\text{s}}\left(\mathtt{x}, \mathtt{y}, \epsilon\right) = \frac{1}{R'\left(1\right)}H\left(R^{\boxtimes}\left(\mathtt{x}\right)\right) + H\left(\mathtt{y}\right) - H\left(\mathtt{x} \boxtimes \mathtt{y}\right) - \frac{1}{L'\left(1\right)}H\left(\mathtt{c}_{\epsilon}\boxdot L^{\boxdot}\left(\mathtt{y}\right)\right).
\end{equation}

Following (\ref{Lemma: Duality Rule for Entropy}), we can rewrite $W_{\text{s}}\left(\mathtt{x}, \mathtt{y}, \epsilon\right)$ as follows
\begin{equation}
\label{Definition: Ws 2}
W_{\text{s}}\left(\mathtt{x}, \mathtt{y}, \epsilon\right) = \frac{1}{R'\left(1\right)}H\left(R^{\boxtimes}\left(\mathtt{x}\right)\right) - H\left(\mathtt{x}\right) + H\left(\mathtt{x} \boxdot \mathtt{y}\right) - \frac{1}{L'\left(1\right)}H\left(\mathtt{c}_{\epsilon}\boxdot L^{\boxdot}\left(\mathtt{y}\right)\right).
\end{equation}

Obviously, the relation between $U_{\text{s}}\left(\mathtt{x}, \epsilon\right)$ and $W_{\text{s}}\left(\mathtt{x}, \mathtt{y}, \epsilon\right)$ is given by
\begin{equation}
U_{\text{s}}\left(\mathtt{x}, \epsilon\right) = \left.\frac{1}{L'\left(1\right)} W_{\text{s}}\left(\mathtt{x}, \mathtt{y}, \epsilon\right)\right|_{\mathtt{y} = \rho^{\boxtimes}\left(\mathtt{x}\right)}.
\end{equation}

Let $\mathtt{x}^{\left(0\right)} = \mathtt{a}$ and $\mathtt{y}^{\left(0\right)} = \rho^{\boxtimes}\left(\mathtt{a}\right)$. By assumption and Claims 3) and 4) in Lemma \ref{LEM: Monotonicity of Ts}, the inequality (\ref{Lemma: Monotonicity of Us}) is equivalent to the fact that the following sequence
\begin{equation}
\left\{W_{\text{s}}\left(\mathtt{x}^{\left(l\right)}, \mathtt{y}^{\left(l\right)}, \epsilon\right)\right\}_{l \in \mathbb{Z}}
\end{equation}
is nonincreasing as $l$ increases.
To prove this fact, consider
\begin{align}
&W_{\text{s}}\left(\mathtt{x}^{\left(l + 1\right)}, \mathtt{y}^{\left(l + 1\right)}, \epsilon\right) - W_{\text{s}}\left(\mathtt{x}^{\left(l\right)}, \mathtt{y}^{\left(l\right)}, \epsilon\right) \nonumber \\
\label{Proof: PotFun is decreasing as l increases}
=& \left[W_{\text{s}}\left(\mathtt{x}^{\left(l + 1\right)}, \mathtt{y}^{\left(l + 1\right)}, \epsilon\right) - W_{\text{s}}\left(\mathtt{x}^{\left(l\right)}, \mathtt{y}^{\left(l + 1\right)}, \epsilon\right)\right] + \left[W_{\text{s}}\left(\mathtt{x}^{\left(l\right)}, \mathtt{y}^{\left(l + 1\right)}, \epsilon\right) - W_{\text{s}}\left(\mathtt{x}^{\left(l\right)}, \mathtt{y}^{\left(l\right)}, \epsilon\right)\right].
\end{align}

We reformulate the term in the first square bracket on the right-hand side of (\ref{Proof: PotFun is decreasing as l increases}) as follows,
\begin{align}
& W_{\text{s}}\left(\mathtt{x}^{\left(l + 1\right)}, \mathtt{y}^{\left(l + 1\right)}, \epsilon\right) - W_{\text{s}}\left(\mathtt{x}^{\left(l\right)}, \mathtt{y}^{\left(l + 1\right)}, \epsilon\right) \nonumber \\
=& \frac{1}{R'\left(1\right)}H\left(R^{\boxtimes}\left(\mathtt{x}^{\left(l + 1\right)}\right) - R^{\boxtimes}\left(\mathtt{x}^{\left(l\right)}\right)\right) -H\left( \left(\mathtt{x}^{\left(l + 1\right)} - \mathtt{x}^{\left(l\right)}\right) \boxtimes \mathtt{y}^{\left(l + 1\right)}\right) \\
=& H\Big(\left(\mathtt{x}^{\left(l + 1\right)} - \mathtt{x}^{\left(l\right)}\right)^{\boxtimes 2} \boxtimes \sum_{j} \frac{\rho_{j}}{j} \sum_{j_0 = 0}^{j - 1} \mathtt{x}^{\left(l\right) \boxtimes j - 1 - j_0} \boxtimes \sum_{j_1 = 0}^{j_0 - 1} \mathtt{x}^{\left(l\right) \boxtimes j_0 - 1 - j_1} \boxtimes \mathtt{x}^{\left(l + 1\right) \boxtimes j_1}\Big).
\end{align}

Similarly, the other term can be rewritten as

\begin{align}
& W_{\text{s}}\left(\mathtt{x}^{\left(l\right)}, \mathtt{y}^{\left(l + 1\right)}, \epsilon\right) - W_{\text{s}}\left(\mathtt{x}^{\left(l\right)}, \mathtt{y}^{\left(l\right)}, \epsilon\right) \nonumber \\
=& H\left(\mathtt{x}^{\left(l\right)} \boxdot \left(\mathtt{y}^{\left(l + 1\right)} - \mathtt{y}^{\left(l\right)}\right)\right) - \frac{1}{L'\left(1\right)}H\left(\mathtt{c} \boxdot L^{\boxdot} \left(\mathtt{y}^{\left(l + 1\right)}\right) - \mathtt{c} \boxdot L^{\boxdot} \left(\mathtt{y}^{\left(l\right)}\right) \right) \\
=& - H\Big(\left(\mathtt{y}^{\left(l + 1\right)} - \mathtt{y}^{\left(l\right)}\right)^{\boxdot 2} \boxdot \mathtt{c} \boxdot \sum_{i} \frac{\lambda_{i}}{i} \sum_{i_{0} = 0}^{i - 1} \mathtt{y}^{\left(l\right) \boxdot i - 1 - i_{0}} \boxdot \sum_{i_{1} = 0}^{i_{0} - 1} \mathtt{y}^{\left(l\right) \boxdot i_{0} - 1 - i_{1}} \boxdot \mathtt{y}^{\left(l + 1\right) \boxdot i_{1}}\Big).
\end{align}

By Corollary \ref{COL: Conv of Square Diff Densities}, the above terms are both nonnegative and hence
\begin{equation}
W_{\text{s}}\left(\mathtt{x}^{\left(l + 1\right)}, \mathtt{y}^{\left(l + 1\right)}, \epsilon\right) \le W_{\text{s}}\left(\mathtt{x}^{\left(l\right)}, \mathtt{y}^{\left(l\right)}, \epsilon\right), \forall l \in \mathbb{Z}.
\end{equation}
This leads to the desired result (\ref{Lemma: Monotonicity of Us}).
\end{IEEEproof}

\begin{definition}
\label{DEF: Density Direction}
A direction defined over $\mathcal{X}$, denoted as $\delta\mathtt{x} = \left(\delta x_0, \delta x_1, \ldots, \delta x_m\right)$, is a vector of length $m + 1$, satisfying $\sum_{i = 0}^{m} \delta x_{i} = 0$.
For convenience, in the sequel, we will always consider $\delta x_1, \ldots, \delta x_m$ as independent variables and rewrite the first entry $\delta x_0$ as $\delta x_0 = - \sum_{i = 1}^{m} \delta x_{i}$.
As a result, whenever we speak of a direction $\delta\mathtt{x}$, we always rewrite it in the form
\begin{equation}
\label{Definition: Density Direction 1}
\delta\mathtt{x} = \left( \left(- \sum_{i = 1}^{m} \delta x_{i}\right), \delta x_1, \ldots, \delta x_m\right),
\end{equation}
or equivalently,
\begin{equation}
\label{Definition: Density Direction 2}
\delta\mathtt{x} = \sum_{i = 1}^{m} \left(\mathtt{\Delta}_{i} - \mathtt{\Delta}_{0}\right) \delta x_i.
\end{equation}
\end{definition}

\begin{definition}
\label{DEF: Directional Derivative}
The directional derivative of $U_{\text{s}}\left(\mathtt{x}, \epsilon\right)$ with respect to $\mathtt{x}$ in the direction $\delta\mathtt{x}$ is defined as
\begin{equation}
\text{d}_{\mathtt{x}}U_{\text{s}}\left(\mathtt{x}, \epsilon\right)\left[\delta\mathtt{x}\right] = \lim_{t \to 0} \frac{U_{\text{s}}\left(\mathtt{x} + t\delta\mathtt{x}, \epsilon\right) - U_{\text{s}}\left(\mathtt{x}, \epsilon\right)}{t}.
\end{equation}
\end{definition}

\begin{lemma}
\label{LEM: Formula for Directional Derivative}
The directional derivative $\text{d}_{\mathtt{x}}U_{\text{s}}\left(\mathtt{x}, \epsilon\right)\left[\delta\mathtt{x}\right]$ defined as above is determined by
\begin{equation}
\text{d}_{\mathtt{x}}U_{\text{s}}\left(\mathtt{x}, \epsilon\right)\left[\delta\mathtt{x}\right] = L'\left(1\right) H\left(\left(\mathtt{T}_\text{s}^{\left(1\right)}\left(\mathtt{x}, \mathtt{c}_{\epsilon}\right) - \mathtt{x}\right)\boxtimes \rho'^{\boxtimes}\left(\mathtt{x}\right)\boxtimes\delta\mathtt{x}\right)
\end{equation}
where $\rho'^{\boxtimes}\left(\mathtt{x}\right)$ is given by
\begin{equation}
\rho'^{\boxtimes}\left(\mathtt{x}\right) = \sum_{j} \left(j - 1\right)\rho_j \mathtt{x}^{\boxtimes {j - 2}}.
\end{equation}
\end{lemma}

The proof of Lemma \ref{LEM: Formula for Directional Derivative} is identical to that of Lemma 23 in \cite{6912949}.
One may also prove this lemma based on the derivative of $U_{\text{s}}\left(\mathtt{x}, \epsilon\right)$ with respect to $\mathtt{x}$.
See Appendix \ref{APP: Three Laws of Basic Operators} for details.

\begin{definition}
\label{DEF: PotFun for StP}
A density $\mathtt{x} \in \mathcal{X}$ is a stationary point of $U_{\text{s}}\left(\mathtt{x}, \epsilon\right)$ if
\begin{equation}
\label{Definition: PotFun for StP}
\text{d}_{\mathtt{x}}U_{\text{s}}\left(\mathtt{x}, \epsilon\right)\left[\delta\mathtt{x}\right] = 0
\end{equation}
for any direction $\delta\mathtt{x}$.
\end{definition}

\begin{definition}
\label{DEF: Energy Gap}
For the uncoupled DE recursion (\ref{Definition: Uncoupled DE 2}), we define the energy gap as
\begin{equation}
\label{Definition: DeltaE}
\Delta E\left(\epsilon\right) = \min_{\mathtt{x} \in \mathcal{F}_{\text{s}}\left(\epsilon\right) \backslash \left\{\mathtt{\Delta}_{0}\right\}} U_{\text{s}}\left(\mathtt{x}, \epsilon\right)
\end{equation}
with the convention that the minimum over the empty set is $ + \infty$.
\end{definition}

\begin{remark}
Notice that in this paper the definition of $\Delta E\left(\epsilon\right)$ is different from that introduced in \cite{6912949}.
While in \cite{6912949} the infimum of $U_{\text{s}}\left(\mathtt{x}, \epsilon\right)$ is over the densities outside the basin of attraction to $\mathtt{\Delta}_{0}$ (see Definition 25 therein), the minimization in (\ref{Definition: DeltaE}) is over the UFP set $\mathcal{F}_{\text{s}}\left(\epsilon\right)$ excluding $\mathtt{\Delta}_{0}$.
This modification is based on the numerical observation that the asymptotic BP threshold of SC-LDPC$\left(\lambda, \rho, N, w, {m}\right)$ on the BEC asymptotically is closely related to the sign of $U_{\text{s}}\left(\mathtt{x}, \epsilon\right)$ at a nontrivial UFP $\mathtt{x} \succ \mathtt{\Delta}_{0}$.
\end{remark}

\begin{lemma}
\label{LEM: Monotonicity of DeltaE}
For any $\epsilon_1, \epsilon_2 \in \left[0, 1\right]$ with $\epsilon_1 > \epsilon_2$, we have

1) $U_{\text{s}}\left(\mathtt{x}, \epsilon_1\right) < U_{\text{s}}\left(\mathtt{x}, \epsilon_2\right)$ if $\mathtt{x} \neq \mathtt{\Delta}_{0}$.

2) $\Delta E\left(\epsilon_1\right) < \Delta E\left(\epsilon_2\right)$.
\end{lemma}

\begin{IEEEproof}
Based on the propositions and lemmas in Subsection \ref{Partial Ordering}, we can prove Claim 1) following the same line as Lemma 26 in \cite{6912949}.

Now we put our focus on Claim 2).
For $\epsilon = \epsilon_{2}$, let $\mathtt{x}_{2}$ be the minimizer of $U_{\text{s}}\left(\mathtt{x}, \epsilon_{2}\right)$ over $\mathcal{F}_{\text{s}} \backslash \left\{\mathtt{\Delta}_{0}\right\}$.
Since $\mathtt{x}_{2}$ is a UFP of the uncoupled DE recursion (\ref{Definition: Uncoupled DE 2}), following Claim 1), we have
\begin{equation}
\label{Proof: Monotonicity of DeltaE 1}
\Delta E\left(\epsilon_{2}\right) = U_{\text{s}}\left(\mathtt{x}_{2}, \epsilon_{2}\right) > U_{\text{s}}\left(\mathtt{x}_{2}, \epsilon_{1}\right).
\end{equation}

Further, since $\mathtt{c}_{\epsilon_{1}} \succ \mathtt{c}_{\epsilon_{2}}$, we have
\begin{equation}
\label{Proof: Monotonicity of DeltaE 1.5}
\mathtt{x}_{2} = \mathtt{T}_\text{s}^{\left(1\right)}\left(\mathtt{x}_{2}, \mathtt{c}_{\epsilon_{2}}\right) \stackrel{(a)}{\succeq} \mathtt{T}_\text{s}^{\left(1\right)}\left(\mathtt{x}_{2}, \mathtt{c}_{\epsilon_{1}}\right)
\end{equation}
where (a) follows from Claim 2) in Lemma \ref{LEM: Monotonicity of Ts}.

The inequality (\ref{Proof: Monotonicity of DeltaE 1.5}) indicates that if we increase the channel erasure probability and set the initial density of the uncoupled DE recursion (\ref{Definition: Uncoupled DE 2}) as $\mathtt{x}^{\left(0\right)} = \mathtt{x}_{2}$, then the densities generated by this recursion are always partially ordered and finally converge to a new UFP, denoted as $\mathtt{x}_{1}$ (see Claim 3) in Lemma \ref{LEM: Monotonicity of Ts}). Further, by Lemma \ref{LEM: Monotonicity of Us} and (\ref{Definition: DeltaE}), we have
\begin{align}
\label{Proof: Monotonicity of DeltaE 2}
U_{\text{s}}\left(\mathtt{x}_{2}, \epsilon_{1}\right) \ge U_{\text{s}}\left(\mathtt{T}_\text{s}^{\left(1\right)}\left(\mathtt{x}_{2}, \mathtt{c}_{\epsilon_{1}}\right), \epsilon_{1}\right) \ge U_{\text{s}}\left(\mathtt{T}_\text{s}^{\left(2\right)}\left(\mathtt{x}_{2}, \mathtt{c}_{\epsilon_{1}}\right), \epsilon_{1}\right) \ge \ldots \ge U_{\text{s}}\left(\mathtt{x}_{1}, \epsilon_{1}\right) \ge \Delta E\left(\epsilon_{1}\right).
\end{align}

Now combining (\ref{Proof: Monotonicity of DeltaE 1}) and (\ref{Proof: Monotonicity of DeltaE 2}) yields $\Delta E\left(\epsilon_1\right) < \Delta E\left(\epsilon_2\right)$.
\end{IEEEproof}

\begin{definition}
We define the potential threshold for the uncoupled DE recursion (\ref{Definition: Uncoupled DE 2}) as
\begin{equation}
\epsilon^{\text{pot}} = \sup\left\{\epsilon \in \left[0, 1\right]\left| \Delta E\left(\epsilon\right) > 0  \right.\right\}.
\end{equation}
\end{definition}

%

%
%
%
%

\subsection{SC-LDPC$\left(\lambda, \rho, N, w, {m}\right)$}
\begin{definition}
For SC-LDPC$\left(\lambda, \rho, N, w, {m}\right)$ on the BEC with erasure probability $\epsilon \in  \left[0, 1\right]$, the coupled DE recursion in the $l$-th iteration is given by
\begin{equation}
\label{Definition: Coupled DE}
\mathtt{x}_{i}^{\left(l\right)} = \frac{1}{w} \sum_{k = 0}^{w - 1} \mathtt{c}_{\epsilon, i - k} \boxdot \lambda^{\boxdot} \left( \frac{1}{w} \sum_{j = 0}^{w - 1} \rho^{\boxtimes} \left(\mathtt{x}_{i - k + j}^{\left(l - 1\right)}\right) \right)
\end{equation}
$\forall l \in \mathbb{Z}_{+}$, where $\mathtt{x}_{i}^{\left(l\right)}$ denotes the input density for the CNs at position $i$ and the respective channel density $\mathtt{c}_{\epsilon, i} = \mathtt{c}$ for $i \in \mathbb{N}_{\text{v}}$ and $\mathtt{c}_{\epsilon, i} = \mathtt{\Delta}_{0}$ otherwise.
\end{definition}

In the sequel, $\forall l \in \mathbb{Z}$, we will use $\underline{\mathtt{x}}^{\left(l\right)}$ to represent a density sequence of length $N_{w}$\footnote{Unless otherwise specified, whenever we speak of a density sequence, we always assume that its length is given by $N_{w}$.}, the $i$-th entry of which is denoted as $\mathtt{x}_{i}^{\left(l\right)}, \forall i \in \mathbb{N}_{\text{c}}$.
The set of all such density sequences is denoted as $\mathcal{X}^{N_{w}}$.
In addition, we will adopt $\underline{\mathtt{\Delta}_{0}} = \left(\mathtt{\Delta}_{0}, \mathtt{\Delta}_{0}, \ldots, \mathtt{\Delta}_{0}\right)$ and $\underline{\mathtt{\Delta}_{m}} = \left(\mathtt{\Delta}_{m}, \mathtt{\Delta}_{m}, \ldots, \mathtt{\Delta}_{m}\right)$ to denote the two extremal density sequences in $\mathcal{X}^{N_{w}}$.
Also, we will use the operator $\mathtt{T}_{\text{c}}^{\left(l\right)}\left(\cdot, \cdot\right)$ to denote the coupled DE recursion (\ref{Definition: Coupled DE}) over $l$ iterations, i.e.,
\begin{equation}
\begin{cases}
\underline{\mathtt{x}}^{\left(1\right)} = \mathtt{T}_{\text{c}}^{\left(1\right)}\left(\underline{\mathtt{x}}^{\left(0\right)}, \mathtt{c}_{\epsilon}\right), \\
\underline{\mathtt{x}}^{\left(2\right)} = \mathtt{T}_{\text{c}}^{\left(1\right)}\left(\underline{\mathtt{x}}^{\left(1\right)}, \mathtt{c}_{\epsilon}\right) = \mathtt{T}_{\text{c}}^{\left(2\right)}\left(\underline{\mathtt{x}}^{\left(0\right)}, \mathtt{c}_{\epsilon}\right), \\
\ldots \\
\underline{\mathtt{x}}^{\left(l\right)} = \mathtt{T}_{\text{c}}^{\left(1\right)}\left(\underline{\mathtt{x}}^{\left(l-1\right)}, \mathtt{c}_{\epsilon}\right) = \mathtt{T}_{\text{c}}^{\left(2\right)}\left(\underline{\mathtt{x}}^{\left(l-2\right)}, \mathtt{c}_{\epsilon}\right) = \ldots = \mathtt{T}_{\text{c}}^{\left(l\right)}\left(\underline{\mathtt{x}}^{\left(0\right)}, \mathtt{c}_{\epsilon}\right).
\end{cases}
\end{equation}

\begin{definition}
For a fixed $\epsilon \in \left[0, 1\right]$, a density sequence $\underline{\mathtt{x}} \in \mathcal{X}^{N_{w}}$ is said to be a coupled fixed point (CFP) of the coupled DE recursion (\ref{Definition: Coupled DE}) if it satisfies $\underline{\mathtt{x}} = \mathtt{T}_\text{c}^{\left(1\right)}\left(\underline{\mathtt{x}}, \mathtt{c}_{\epsilon}\right)$.
In the sequel, we will use $\mathcal{F}_{\text{c}}\left(\epsilon, N, w\right)$ to denote the set of all such CFPs.
\end{definition}

We define partial ordering between density sequences in a pointwise manner, i.e., for any $\underline{\mathtt{x}}, \underline{\mathtt{y}} \in \mathcal{X}^{N_{w}}$, we say that $\underline{\mathtt{x}} \preceq \underline{\mathtt{y}}$ or $\underline{\mathtt{y}} \succeq \underline{\mathtt{x}}$ if $\mathtt{x}_{i} \preceq \mathtt{y}_{i}, \forall i \in \mathbb{N}_{\text{c}}$.
Further, we say that $\underline{\mathtt{x}} \prec \underline{\mathtt{y}}$ or $\underline{\mathtt{y}} \succ \underline{\mathtt{x}}$ if $\mathtt{x}_{i} \prec \mathtt{y}_{i}, \forall i \in \mathbb{N}_{\text{c}}$.


\begin{lemma}
\label{LEM: Monotonicity of Tc}
Consider $\underline{\mathtt{a}},\underline{\mathtt{a}}_1, \underline{\mathtt{a}}_2 \in \mathcal{X}^{N_{w}}$ and $\mathtt{c}, \mathtt{c}_1, \mathtt{c}_2 \in \mathcal{X}$. For the coupled DE recursion (\ref{Definition: Coupled DE}) with $l \in \mathbb{Z}_{+}$, we have

1) If $\underline{\mathtt{a}}_1 \succeq \underline{\mathtt{a}}_2$, then $\mathtt{T}_{\text{c}}^{\left(l\right)}\left(\underline{\mathtt{a}}_1, \mathtt{c}\right) \succeq \mathtt{T}_{\text{c}}^{\left(l\right)}\left(\underline{\mathtt{a}}_2, \mathtt{c}\right), \forall \mathtt{c} \in \mathcal{X}$.

2) If $\mathtt{c}_1 \succeq \mathtt{c}_2$, then $\mathtt{T}_{\text{c}}^{\left(l\right)}\left(\underline{\mathtt{a}}, \mathtt{c}_1\right) \succeq \mathtt{T}_{\text{c}}^{\left(l\right)}\left(\underline{\mathtt{a}}, \mathtt{c}_2\right), \forall \underline{\mathtt{a}} \in \mathcal{X}^{N_{w}}$.

3) If $\exists \underline{\mathtt{a}} \in \mathcal{X}^{N_{w}}$ such that $\mathtt{T}_{\text{c}}^{\left(1\right)}\left(\underline{\mathtt{a}}, \mathtt{c}\right) \preceq \underline{\mathtt{a}}$, then $\mathtt{T}_{\text{c}}^{\left(l + 1\right)}\left(\underline{\mathtt{a}}, \mathtt{c}\right) \preceq \mathtt{T}_{\text{c}}^{\left(l\right)}\left(\underline{\mathtt{a}}, \mathtt{c}\right)$ and the limit $\mathtt{T}_{\text{c}}^{\left(\infty\right)}\left(\underline{\mathtt{a}}, \mathtt{c}\right) = \lim_{l \to \infty} \mathtt{T}_{\text{c}}^{\left(l\right)}\left(\underline{\mathtt{a}}, \mathtt{c}\right)$ does indeed exist, satisfying $\mathtt{T}_{\text{c}}^{\left(\infty\right)}\left(\underline{\mathtt{a}}, \mathtt{c}\right) \preceq \mathtt{T}_{\text{c}}^{\left(l\right)}\left(\underline{\mathtt{a}}, \mathtt{c}\right)$ and
\begin{equation}
\mathtt{T}_{\text{c}}^{\left(1\right)}\left(\mathtt{T}_{\text{c}}^{\left(\infty\right)}\left(\underline{\mathtt{a}}, \mathtt{c}\right), \mathtt{c}\right)  = \mathtt{T}_{\text{c}}^{\left(\infty\right)}\left(\underline{\mathtt{a}}, \mathtt{c}\right).
\end{equation}

4) If $\exists \underline{\mathtt{a}} \in \mathcal{X}^{N_{w}}$ such that $\mathtt{T}_{\text{c}}^{\left(1\right)}\left(\underline{\mathtt{a}}, \mathtt{c}\right) \succeq \underline{\mathtt{a}}$, then $\mathtt{T}_{\text{c}}^{\left(l + 1\right)}\left(\underline{\mathtt{a}}, \mathtt{c}\right) \succeq \mathtt{T}_{\text{c}}^{\left(l\right)}\left(\underline{\mathtt{a}}, \mathtt{c}\right)$ and the limit $\mathtt{T}_{\text{c}}^{\left(\infty\right)}\left(\underline{\mathtt{a}}, \mathtt{c}\right) = \lim_{l \to \infty} \mathtt{T}_{\text{c}}^{\left(l\right)}\left(\underline{\mathtt{a}}, \mathtt{c}\right)$ does indeed exist, satisfying $\mathtt{T}_{\text{c}}^{\left(\infty\right)}\left(\underline{\mathtt{a}}, \mathtt{c}\right) \succeq \mathtt{T}_{\text{c}}^{\left(l\right)}\left(\underline{\mathtt{a}}, \mathtt{c}\right)$ and
\begin{equation}
\mathtt{T}_{\text{c}}^{\left(1\right)}\left(\mathtt{T}_{\text{c}}^{\left(\infty\right)}\left(\underline{\mathtt{a}}, \mathtt{c}\right), \mathtt{c}\right)  = \mathtt{T}_{\text{c}}^{\left(\infty\right)}\left(\underline{\mathtt{a}}, \mathtt{c}\right).
\end{equation}
\end{lemma}
\begin{IEEEproof}
See the proof of Lemma 34 in \cite{6912949}.
\end{IEEEproof}

For brevity, in the sequel, unless otherwise specified, whenever we speak of a coupled DE recursion, we always assume that $\underline{\mathtt{x}}^{\left(0\right)} = \underline{\mathtt{\Delta}_{m}}$.
Under this assumption, Claim 3) in Lemma \ref{LEM: Monotonicity of Tc} indicates that the density sequences generated by this recursion are always partially ordered.
Further, these sequences satisfy the following symmetric constraint due to the uniform coupling weights and symmetric boundary conditions \cite{6912949},
\begin{equation}
\label{None: Symmetry of a CFP}
{\mathtt{x}}_{i}^{\left(l\right)} = {\mathtt{x}}_{N_{w} - i}^{\left(l\right)}, \forall i \in \mathbb{N}_{\text{c}}.
\end{equation}

Due to the above constraint, we focus our discussion on the ``middle point'' of a CFP $\underline{\mathtt{x}} \in \mathcal{F}_{\text{c}}\left(\epsilon, N, w\right)$, i.e., ${\mathtt{x}}_{N^{\text{mid}}_{w}}$. For a fixed $\epsilon \in \left[0, 1\right]$, we write ${\mathtt{x}}_{N^{\text{mid}}_{w}}$ in the form of ${\mathtt{x}}_{N^{\text{mid}}_{w}} = {\mathtt{m}} \left(N, w\right)$ to highlight the fact that this density depends on the coupling length $N$ and the coupling width $w$.
By doing this, we can show that ${\mathtt{m}} \left(N, w\right)$ converges to a UFP as $N \to \infty$, as stated in the following lemma.

\begin{lemma}
\label{LEM: The Limit of the Middle Point}
For any fixed $\epsilon \in \left[0, 1\right]$ and $w \in \mathbb{Z}_{+}$, the limit ${\mathtt{m}}\left(\infty, w\right) = \lim_{N \to \infty} {\mathtt{m}} \left(N, w\right)$ exists.
Further, it is a UFP of the uncoupled DE recursion (\ref{Definition: Uncoupled DE 2}), i.e., ${\mathtt{m}}\left(\infty, w\right) \in \mathcal{F}_{\text{s}}\left(\epsilon\right)$.
\end{lemma}

\begin{IEEEproof}
Now consider two coupled DE recursions sharing the same degree distribution pair $\left(\lambda, \rho\right)$, coupling width $w$ and channel erasure probability $\epsilon$, but with different coupling lengths $N'$ and $N$ where $N' < N$.
Denote by $\underline{\mathtt{x}}'$ and $\underline{\mathtt{x}}$ the CFPs of these two coupled DE recursions.

Following Lemma \ref{LEM: Monotonicity of Tc}, it is easy to verify the following facts:

1) ${\mathtt{x}'}_{i} \preceq {\mathtt{x}}_{i}$.

2) ${\mathtt{x}'}_{i} \preceq {\mathtt{x}'}_{i + 1}$, $\forall i \in \left\{0, 1, \ldots, \lfloor \left(N' + w - 1\right)/2 \rfloor -1\right\}$.

3) ${\mathtt{x}}_{i} \preceq {\mathtt{x}}_{i + 1}$, $\forall i \in \left\{0, 1, \ldots, \lfloor \left(N + w - 1\right)/2 \rfloor -1\right\}$.

Further, we can conclude from the above facts that
\begin{equation}
\label{Proof: The Middle Point of a CFP}
{\mathtt{x}'}_{\lfloor \left(N' + w - 1\right)/2 \rfloor} \preceq {\mathtt{x}}_{\lfloor \left(N' + w - 1\right)/2 \rfloor} \preceq {\mathtt{x}}_{\lfloor \left(N + w - 1\right)/2 \rfloor}.
\end{equation}

Therefore, we have ${\mathtt{m}} \left(N', w\right) \preceq {\mathtt{m}} \left(N, w\right)$, and by Proposition \ref{PRP: Convergence of Partially Ordered Densitiy}, the limit $\lim_{N \to \infty} {\mathtt{m}} \left(N, w\right)$ indeed exists.

Now by fixing $N = N' + 4w$ and letting $N \to \infty$ (thereby $N' \to \infty$), we can rewrite the inequality (\ref{Proof: The Middle Point of a CFP}) as\footnote{In general, if a density sequence $\underline{\mathtt{x}}$ is a CFP of the coupled DE recursion (\ref{Definition: Coupled DE}), then each entry of $\underline{\mathtt{x}}$ implicitly depends on the coupling length $N$.}
\begin{equation}
{\mathtt{m}}\left(\infty, w\right) = \lim_{N' \to \infty}{\mathtt{x}'}_{\lfloor \left(N' + w - 1\right)/2 \rfloor} \preceq \lim_{N \to \infty} {\mathtt{x}}_{\lfloor \left(N' + w - 1\right)/2 \rfloor} \preceq \lim_{N \to \infty} {\mathtt{x}}_{\lfloor \left(N + w - 1\right)/2 \rfloor} = {\mathtt{m}}\left(\infty, w\right) .
\end{equation}

Therefore, $\forall k \in \left\{0, 1, \ldots, 2w\right\}$,
\begin{equation}
\lim_{N \to \infty} {\mathtt{x}}_{\lfloor \left(N + w - 1\right)/2 \rfloor -k} = {\mathtt{m}}\left(\infty, w\right).
\end{equation}

By substituting the above limit to the following CFP equation with $i = \lfloor \left(N + w - 1\right)/2 \rfloor - w$,
\begin{equation}
\label{Definition: CFP of Coupled DE}
\mathtt{x}_{i} = \frac{1}{w} \sum_{k = 0}^{w - 1} \mathtt{c}_{\epsilon, i - k} \boxdot \lambda^{\boxdot} \left( \frac{1}{w} \sum_{j = 0}^{w - 1} \rho^{\boxtimes} \left(\mathtt{x}_{i - k + j}\right) \right),
\end{equation}
we obtain the following UFP equation ${\mathtt{m}}\left(\infty, w\right) = \mathtt{c} \boxdot \lambda^{\boxdot} \left(  \rho^{\boxtimes} \left({\mathtt{m}}\left(\infty, w\right)\right) \right)$, i.e., ${\mathtt{m}}\left(\infty, w\right)$ is a UFP of the uncoupled DE recursion (\ref{Definition: Uncoupled DE 2}).
\end{IEEEproof}

\begin{definition}
For any $\underline{\mathtt{x}} \in \mathcal{X}^{N_{w}}, \epsilon \in \left[0, 1\right]$, the potential function for SC-LDPC$\left(\lambda, \rho, N, w, {m}\right)$ is given by
\begin{equation}
U_{\text{c}}\left(\underline{\mathtt{x}}, \epsilon\right) = L'\left(1\right) \sum_{i = 0}^{N^{\text{mid}}_{w}}H\Bigg(\frac{1}{R'\left(1\right)}R^{\boxtimes}\left(\mathtt{x}_{i}\right) + \rho^{\boxtimes}\left(\mathtt{x}_{i}\right)  - \mathtt{x}_{i} \boxtimes \rho^{\boxtimes}\left(\mathtt{x}_{i}\right) - \mathtt{c} \boxdot L^{\boxdot}\bigg(\frac{1}{w} \sum_{j = 0}^{w - 1} \rho^{\boxtimes}\left(\mathtt{x}_{i + j}\right)\bigg) \Bigg).
\end{equation}
\end{definition}

\begin{remark}
Notice that the potential function $U_{\text{c}}\left(\underline{\mathtt{x}}, \epsilon\right)$ defined in this paper is slightly different from \cite{6912949} (see Definition 37 therein),
Here we restrict the sum over $i \in \left\{0, 1, \ldots, N^{\text{mid}}_{w}\right\}$ based on the symmetric constraint (\ref{None: Symmetry of a CFP}), regarding the entries of the former half of $\underline{\mathtt{x}}$ as independent variables.
Due to the same reason, we define the direction over $\mathcal{X}^{N_{w}}$ as follows.
\end{remark}

\begin{definition}
A direction over $\mathcal{X}^{N_{w}}$, denoted as $\underline{\delta\mathtt{x}}$, is a sequence of length $N_{w}$, the first $N^{\text{mid}}_{w} + 1$ entries of which are independent directions defined over $\mathcal{X}$ and the others are zero vectors of length $m + 1$, i.e.,
\begin{equation}
\underline{\delta\mathtt{x}} = \left(\delta\mathtt{x}_0, \delta\mathtt{x}_1, \ldots, \delta\mathtt{x}_{N^{\text{mid}}_{w}}, \mathtt{0}, \ldots, \mathtt{0}\right).
\end{equation}
\end{definition}

\begin{definition}
The directional derivative of $U_{\text{c}}\left(\underline{\mathtt{x}}, \epsilon\right)$ with respect to $\underline{\mathtt{x}}$ in the direction $\underline{\delta\mathtt{x}}$ is defined as
\begin{equation}
\text{d}_{\underline{\mathtt{x}}}U_{\text{c}}\left(\underline{\mathtt{x}}, \epsilon\right)\left[\underline{\delta\mathtt{x}}\right] = \lim_{t \to 0} \frac{U_{\text{c}}\left(\underline{\mathtt{x}} + t\underline{\delta\mathtt{x}}, \epsilon\right) - U_{\text{c}}\left(\underline{\mathtt{x}}, \epsilon\right)}{t}.
\end{equation}
\end{definition}

\begin{lemma}
\label{LEM: Derivative of Uc}
The directional derivative of $U_{\text{c}}\left(\underline{\mathtt{x}}, \epsilon\right)$ defined as above is given by
\begin{equation}
\text{d}_{\underline{\mathtt{x}}}U_{\text{c}}\left(\underline{\mathtt{x}}, \epsilon\right)\left[\underline{\delta\mathtt{x}}\right] = L'\left(1\right) \sum_{i = 0}^{N^{\text{mid}}_{w}} H\left(\left(\left[\mathtt{T}_{\text{c}}^{\left(1\right)}\left(\underline{\mathtt{x}}, \mathtt{c}_{\epsilon}\right) - \underline{\mathtt{x}}\right]_{i} \boxtimes \rho'^{\boxtimes}\left(\mathtt{x}_{i}\right) \boxtimes \delta\mathtt{x}_{i}\right)\right).
\end{equation}
\end{lemma}
\begin{IEEEproof}
The proof of Lemma \ref{LEM: Derivative of Uc} is almost identical to that of Lemma 38 in \cite{6912949}, and we omit the details for brevity.
\end{IEEEproof}

Lemma \ref{LEM: Derivative of Uc} indicates that $\text{d}_{\underline{\mathtt{x}}}U_{\text{c}}\left(\underline{\mathtt{x}}, \epsilon\right)\left[\underline{\delta\mathtt{x}}\right]$ vanishes if $\underline{\mathtt{x}}$ is a CFP of the coupled DE recursion (\ref{Definition: Coupled DE}).

\begin{lemma}
\label{LEM: Change of Uc via Shifting}
Define the shift operator $\mathtt{S}\left(\cdot\right)$ as follows \cite{6912949}
\begin{equation}
\left(\mathtt{S}\left(\underline{\mathtt{x}}\right)\right)_{i} =
\begin{cases}
\mathtt{\Delta}_{0}, &i = 0 \\
\mathtt{x}_{i - 1}, &i \in \mathbb{N}_{\text{c}} \backslash \left\{0\right\}. \\
\end{cases}
\end{equation}
Let $\underline{\mathtt{x}}$ be a CFP of the coupled DE recursion (\ref{Definition: Coupled DE}).
For a fixed $\epsilon \in \left[0, 1\right]$ and an arbitrary small $\eta > 0$, there exists $N_{\eta} \in \mathbb{Z}_{+}$ such that $\forall N > N_{\eta}$, after applying the operator $\mathtt{S}\left(\cdot\right)$ to $\underline{\mathtt{x}}$, the change of $U_{\text{c}}\left(\underline{\mathtt{x}}, \epsilon\right)$ is bounded as follows
\begin{equation}
U_{\text{c}}\left(\mathtt{S}\left(\underline{\mathtt{x}}\right), \epsilon\right) - U_{\text{c}}\left(\underline{\mathtt{x}}, \epsilon\right) < -\Delta E\left(\epsilon\right) + \eta.
\end{equation}
\end{lemma}
\begin{IEEEproof}
First of all, following the same line as in the proof of Lemma 41 in \cite{6912949}, we can show that the change of $U_{\text{c}}\left(\underline{\mathtt{x}}, \epsilon\right)$ is bounded by the underlying potential function at the ``middle point'' of $\underline{\mathtt{x}}$, i.e.,
\begin{equation}
U_{\text{c}}\left(\mathtt{S}\left(\underline{\mathtt{x}}\right), \epsilon\right) - U_{\text{c}}\left(\underline{\mathtt{x}}, \epsilon\right) \le - U_{\text{s}}\left(\mathtt{x}_{N^{\text{mid}}_{w}}, \epsilon\right).
\end{equation}

Next, since the ``middle point'' $\mathtt{x}_{N^{\text{mid}}_{w}}$ converges to a UFP ${\mathtt{m}}\left(\infty, w\right)$ (see Lemma \ref{LEM: The Limit of the Middle Point}), we can deduce from the continuity of $U_{\text{s}}\left(\mathtt{x}, \epsilon\right)$ with respect to $\mathtt{x}$ that for any arbitrary small $\eta > 0$ there exists an integer $N_{\eta}$ such that $\forall N > N_{\eta}, \left|U_{\text{s}}\left(\mathtt{x}_{N^{\text{mid}}_{w}}, \epsilon\right) - U_{\text{s}}\left({\mathtt{m}}\left(\infty, w\right), \epsilon\right)\right| < \eta.$. Therefore,
\begin{equation}
U_{\text{c}}\left(\mathtt{S}\left(\underline{\mathtt{x}}\right), \epsilon\right) - U_{\text{c}}\left(\underline{\mathtt{x}}, \epsilon\right) \le - U_{\text{s}}\left(\mathtt{x}_{N^{\text{mid}}_{w}}, \epsilon\right) < -U_{\text{s}}\left({\mathtt{m}}\left(\infty, w\right), \epsilon\right) + \eta \le -\Delta E\left(\epsilon\right) + \eta.
\end{equation}
\end{IEEEproof}

\subsection{Theorems for Threshold Saturation}

Based on the above propositions and lemmas, we can follow a similar procedure as in \cite{6912949} to establish the following theorem. See the proof of Theorem 44 therein and we do not reproduce the details in this paper.

\begin{theorem}
\label{THR: Threshold Saturation}
Consider an SC-LDPC$\left(\lambda, \rho, N, w, {m}\right)$ ensemble on the BEC with erasure probability $\epsilon \in \left[\left.0, \epsilon^{pot}\right)\right.$.
For arbitrary small $\eta > 0$, there exists $N_{\eta} \in \mathbb{Z}_{+}$, and a positive constant independent of $N$ and $w$, denoted as $K_{\lambda, \rho}$, such that $\forall N > N_{\eta}, w > K_{\lambda, \rho}/\left({\Delta E\left(\epsilon\right)} - \eta \right)$, the only CFP of the coupled DE recursion (\ref{Definition: Coupled DE}) is $\underline{\mathtt{\Delta}_{0}}$.
\end{theorem}

Likewise, the converse to Theorem \ref{THR: Threshold Saturation} can be shown following almost the same line as in the proof of Theorem 47 in \cite{6912949}.

\begin{theorem}
\label{THR: Converse to Threshold Saturation}
Consider an SC-LDPC$\left(\lambda, \rho, N, w_0, m\right)$ ensemble on the BEC with erasure probability $\epsilon \in \left.\left(\epsilon^{\text{pot}}, 1\right.\right]$.
There exists $N_{0} \in \mathbb{Z}_{+}$ such that $\forall N > N_{0}$, the CFP of the coupled DE recursion (\ref{Definition: Coupled DE}) initialized with $\underline{\mathtt{\Delta}_{m}}$ satisfies
\begin{equation}
\mathtt{T}_{\text{c}}^{\left(\infty\right)}\left(\underline{\mathtt{\Delta}_{m}}, \mathtt{c}_{\epsilon}\right) \succ \underline{\mathtt{\Delta}_{0}}.
\end{equation}
\end{theorem}

\section{Conclusion}
\label{Conclusion}
We investigated the asymptotic performance for SC-LDPC ensembles defined over GL$\left(2^{m}\right)$.
Our purpose is to prove the existence of the threshold saturation effect for transmission on the BEC.
To this end, we presented a detailed analysis of the entropy function and the VN and CN convolutional operators and discussed their properties through several propositions and lemmas.
In particular, we derived a nonbinary version of the duality rule for entropy to accommodate the DE analysis of nonbinary LDPC ensembles on the BEC.
Based on this, we constructed potential functions for the uncoupled and coupled DE recursions, the forms of which are very similar to those in \cite{6912949}.
These findings led us to establish the threshold saturation theorem and its converse following almost the same approach developed by S. Kumar \textit{et al}.

\appendices

\section{}
\label{APP: A Useful Tool}
The following proposition is useful in the proofs of some propositions and lemmas in this paper.

\begin{proposition}
\label{PRP: An important TRICK}
Consider two vectors $\left(u_1, u_2, \ldots, u_{K} \right)$ and $\left(v_1, v_2, \ldots, v_{K} \right)$ with $K \in \mathbb{Z}_{+}$.
The following identity holds for $n = 1, \ldots, K-1$,
\begin{equation}
\sum_{i = n}^{K} v_i u_i = v_n \sum_{k = n}^{K} u_k + \sum_{i = n+1}^{K} \left(v_i - v_{i-1}\right) \sum_{k = i}^{K} u_k.
\end{equation}
\end{proposition}

\section{Some Properties of $V_{i, j, k}^{m}$ and $C_{i, j, k}^{m}$}
\label{APP: Some Useful Properties of Basic Operators}
In this Section, we discuss and prove several useful results for $V_{i, j, k}^{m}$ and $C_{i, j, k}^{m}$.

\begin{proposition}
\label{PRP: Properties of V and C}
For any $m \in \mathbb{Z}_{+}$, the coefficients $V_{i, j, k}^{m}$ and $C_{i, j, k}^{m}$ satisfy the following properties.

1) For any $i, j \in \mathbb{M}$, we have $0 \le V_{i, j, k}^{m} \le 1$, $0 \le C_{i, j, k}^{m} \le 1$ and
\begin{equation}
\label{Prp: Sum of V and C over k}
\sum_{k = 0}^{m} V_{i, j, k}^{m} = \sum_{k = 0}^{m} C_{i, j, k}^{m} = 1.
\end{equation}

2) The coefficients $V_{i, j, k}^{m}$ and $C_{i, j, k}^{m}$ remain invariant under a swap of $i$ and $j$, i.e.,
\begin{equation}
\label{Lemma: Symmetry of V and C over ij}
V_{i, j, k}^{m} = V_{j, i, k}^{m}, \quad C_{i, j, k}^{m} = C_{j, i, k}^{m}.
\end{equation}

3) We have $V_{i - 1, j, k}^{m} < V_{i, j, k}^{m}$ if $0 < k \le i \le m$, $0 < k \le j \le m$, and $C_{i - 1, j, k}^{m} > C_{i, j, k}^{m}$ if $0 < i \le k \le m$, $0 < j \le k \le m$.

\end{proposition}

\begin{IEEEproof}
1) The proof of $V_{i, j, k}^{m} \ge 0$ and $C_{i, j, k}^{m} \ge 0$ is trivial by definition. The identities $\sum_{k = 0}^{m} C_{i, j, k}^{m} = \sum\nolimits_{k = 0}^{m} V_{i, j, k}^{m} = 1$ simply follow from the fact that $V_{i, j, k}^{m}$ and $C_{i, j, k}^{m}$ are probabilities (see Subsection II-A in \cite{7430313}).
Alternatively, one may also prove them using the following well-known Vandermonde identity for the $q$-binomial coefficients \cite{kac2001quantum},
\begin{equation}
\label{Proof:LEM: Sum of V and C over k:Eq1}
\genfrac{[}{]}{0pt}{0}{m}{j} = \sum_{k = 0}^{m} 2^{(i - k)(j - k)} \genfrac{[}{]}{0pt}{0}{i}{k} \genfrac{[}{]}{0pt}{0}{m - i}{j - k}.
\end{equation}

2) For $n \in \mathbb{Z}$, define $\left[n\right]$ as follows
\begin{equation}
\label{Definition: [n]}
\left[n\right] =
\begin{cases}
 1, \quad & n = 0 \\
 \prod_{l = 1}^{n} (2^{l} - 1), \quad & \text{otherwise}.
\end{cases}
\end{equation}
The Gaussian binomial coefficient $\genfrac{[}{]}{0pt}{1}{m}{k}$ can be rewritten as
\begin{equation}
\genfrac{[}{]}{0pt}{0}{m}{k} = \frac{\left[m\right]}{\left[m-k\right]\left[k\right]}.
\end{equation}
Rewrite those Gaussian binomial coefficients in (\ref{Definition: Vijk and Cijk}) in the form as above,
\begin{equation}
V_{i, j, k}^{m} = \frac{2^{(i - k)(j - k)} \left[i\right]\left[j\right]\left[m-i\right]\left[m-j\right]}{\left[k\right]\left[m\right]\left[i-k\right]\left[j-k\right]\left[m-i-j+k\right]} = V_{j, i, k}^{m}.
\end{equation}

Similarly, we can show that $C_{i, j, k}^{m} = C_{j, i, k}^{m}$.

3) We focus on the first inequality and omit the proof of the other since $C_{i, j, k}^{m} = V_{m - i, m - j, m - k}^{m}$.
In the case of $0 < k = i \le m$, the first inequality holds since $V_{k - 1, j, k}^{m} = 0 < V_{k, j, k}^{m}$.
For $0 < k < i \le m$, this inequality follows from the fact that $V_{i, j, k}^{m} > 2^{k - 1} V_{i - 1, j, k}^{m}$ (see Appendix A in \cite{7430313}).
\end{IEEEproof}

\section{The commutative, distributive and associative laws of $\boxdot$ and $\boxtimes$}
\label{APP: Three Laws of Basic Operators}
In this section, we aim to prove three important laws of the convolutional operators $\boxdot$ and $\boxtimes$.

\begin{proposition}
Considering three vectors of length $m + 1$ denoted as $\mathtt{a}$, $\mathtt{b}$ and $\mathtt{c}$, we have

1) $\mathtt{a} * \mathtt{b} = \mathtt{b} * \mathtt{a}$.

2) $\mathtt{a} * \left(\mathtt{b} + \mathtt{c}\right) = \mathtt{a} * \mathtt{b} + \mathtt{a} * \mathtt{c}$.

3) $\left(\mathtt{a} * \mathtt{b}\right) * \mathtt{c} = \mathtt{a} * \left(\mathtt{b} * \mathtt{c}\right)$.

\end{proposition}

\begin{IEEEproof}
Claim 1) follows from (\ref{Lemma: Symmetry of V and C over ij}) and Claim 2) can be easily verified by definition.
Thus, we put our focus on Claim 3) for $\boxdot$.
The proof for $\boxtimes$ is identical.

We first compare the $k$-th entries of $\left(\mathtt{a} \boxdot \mathtt{b}\right) \boxdot \mathtt{c}$ and $\mathtt{a} \boxdot \left(\mathtt{b} \boxdot \mathtt{c}\right)$ for any $k \in \mathbb{M}$. On one hand,
\begin{equation}
\left[\left(\mathtt{a} \boxdot \mathtt{b}\right)\boxdot \mathtt{c}\right]_{k} = \sum_{j=0}^{m}\sum_{n=0}^{m} \left[\mathtt{a} \boxdot \mathtt{b}\right]_{j} V_{j, n, k}^{m} c_{n} = \sum_{i=0}^{m}\sum_{l=0}^{m}\sum_{n=0}^{m} a_{i} b_{l} c_{n} \sum_{j=0}^{m} V_{j, n, k}^{m} V_{i, l, j}^{m}.
\end{equation}

On the other hand,
\begin{equation}
\left[\mathtt{a} \boxdot \left(\mathtt{b} \boxdot \mathtt{c}\right)\right]_{k} = \sum_{i=0}^{m}\sum_{j=0}^{m} a_{i} V_{i, j, k}^{m} \left[\mathtt{b} \boxdot \mathtt{c}\right]_{j} = \sum_{i=0}^{m}\sum_{l=0}^{m}\sum_{n=0}^{m} a_{i} b_{l} c_{n} \sum_{j=0}^{m} V_{j, i, k}^{m} V_{n, l, j}^{m}.
\end{equation}
Therefore, $\left(\mathtt{a} \boxdot \mathtt{b}\right)\boxdot \mathtt{c} = \mathtt{a} \boxdot \left(\mathtt{b} \boxdot \mathtt{c}\right)$ holds if
\begin{equation}
\label{Proof: Sum VV = Sum VV}
\sum_{j=0}^{m} V_{j, n, k}^{m} V_{i, l, j}^{m} = \sum_{j=0}^{m} V_{j, i, k}^{m} V_{n, l, j}^{m}.
\end{equation}

In other words, what we need to prove is that either side of (\ref{Proof: Sum VV = Sum VV}) remains invariant when we swap the roles of $i$ and $n$.
To this end, we substitute (\ref{Definition: Vijk and Cijk}) into (\ref{Proof: Sum VV = Sum VV}),
\begin{equation}
\label{Proof: Sum VV LHS}
\sum_{j=0}^{m} V_{j, n, k}^{m} V_{i, l, j}^{m} = \frac{1}{\genfrac{[}{]}{0pt}{0}{m}{i}\genfrac{[}{]}{0pt}{0}{m}{n}} \sum_{j=0}^{m} 2^{(j - k)(i - k) + (l - j)(n - j)} \genfrac{[}{]}{0pt}{0}{j}{k} \genfrac{[}{]}{0pt}{0}{m - j}{i - k}  \genfrac{[}{]}{0pt}{0}{l}{j} \genfrac{[}{]}{0pt}{0}{m - l}{n - j}.
\end{equation}

Applying (\ref{Proof:LEM: Sum of V and C over k:Eq1}) to $\genfrac{[}{]}{0pt}{1}{m - j}{i - k}$, we have
\begin{equation}
\label{Proof: Detaching a GBC}
\genfrac{[}{]}{0pt}{0}{m - j}{i - k} = \genfrac{[}{]}{0pt}{0}{l - j + m - l}{j' - k + i - j'} = \sum_{j'=0}^{m} 2^{\left(i - j'\right)\left(l - j - j' + k\right)}\genfrac{[}{]}{0pt}{0}{l - j}{j' - k}\genfrac{[}{]}{0pt}{0}{ m - l}{i - j'}.
\end{equation}

Substituting (\ref{Proof: Detaching a GBC}) into (\ref{Proof: Sum VV LHS}), we can obtain
\begin{align}
&\sum_{j=0}^{m} V_{j, n, k}^{m} V_{i, l, j}^{m} = \frac{1}{\genfrac{[}{]}{0pt}{0}{m}{i}\genfrac{[}{]}{0pt}{0}{m}{n}}  \sum_{j=0}^{m}\sum_{j'=0}^{m} \genfrac{[}{]}{0pt}{0}{m - l}{n - j} \genfrac{[}{]}{0pt}{0}{m - l}{i - j'} \frac{\left[l\right]}{\left[k\right]\left[j-k\right]\left[j'-k\right]\left[l+k-j-j'\right]} \nonumber \\
&\quad \quad \quad \quad \quad \quad \quad \quad \quad \quad \quad \quad \times 2^{j^2 + jj' + j'^2 - \left(j + j'\right)\left(l + k\right) + k^2 + \left(i + n\right)l - \left(ij' + nj\right)}.
\end{align}

Obviously, swapping the roles of $i$ and $n$ does not change $\sum_{j=0}^{m} V_{j, n, k}^{m} V_{i, l, j}^{m}$, which completes the proof of Clam 3).
\end{IEEEproof}

In the remainder of this appendix, we demonstrate how to apply the above laws to the derivative analysis of the entropy function involving the convolutional operators $\boxdot$ and $\boxtimes$.
For convenience, we write a density $\mathtt{x} \in \mathcal{X}$ in the form of $\mathtt{x} = \left(1 - \sum_{i = 1}^{m}x_{i}, x_1, x_2, \ldots x_{m}\right)$ by regarding $x_1, x_2, \ldots x_m$ as independent variables.
As a result, $\forall i \in \mathbb{M}\backslash \left\{0\right\}$, the partial derivative of $\mathtt{x}$ with respect to $x_{i}$ is a vector of length $m+1$ given by
\begin{equation}
\label{Proof: Derivative of x with xi}
\frac{\partial }{\partial x_{i}}\mathtt{x} = \mathtt{\Delta}_{i} - \mathtt{\Delta}_{0}.
\end{equation}

Therefore, $\forall \mathtt{a} \in \mathcal{X}, i \in \mathbb{M}\backslash \left\{0\right\}$, we have
\begin{equation}
\label{Proof: Derivative of aVx/aCx with xi}
\frac{\partial }{\partial x_{i}}  \left(\mathtt{a} \boxdot \mathtt{x}\right) = \mathtt{a} \boxdot \frac{\partial }{\partial x_{i}}\mathtt{x} \stackrel{(a)}{=} \mathtt{a} \boxdot \mathtt{\Delta}_{i} - \mathtt{\Delta}_{0}, \quad \frac{\partial }{\partial x_{i}}  \left(\mathtt{a} \boxtimes \mathtt{x}\right) = \mathtt{a} \boxtimes \frac{\partial }{\partial x_{i}}\mathtt{x} \stackrel{(b)}{=} \mathtt{a} \boxtimes \mathtt{\Delta}_{i} - \mathtt{a}
\end{equation}
where (a) and (b) are both based on the distributive law of $\boxdot$ and $\boxtimes$.

Moreover, $\forall n \in \mathbb{Z}_{+}, i \in \mathbb{M}\backslash \left\{0\right\}$,
\begin{align}
\label{Proof: Partial deritivative of Exponential *}
\frac{\partial }{\partial x_{i}}\mathtt{x}^{* n} =& \left(\frac{\partial \mathtt{x}}{\partial x_{i}} * \mathtt{x} * \ldots * \mathtt{x}\right) + \left(\mathtt{x} * \frac{\partial \mathtt{x}}{\partial x_{i}} * \ldots * \mathtt{x}\right) + \ldots + \left(\mathtt{x} * \mathtt{x} * \ldots * \frac{\partial \mathtt{x}}{\partial x_{i}}\right) \nonumber \\
\stackrel{(a)}{=}& n \mathtt{x}^{* n - 1} * \frac{\partial\mathtt{x} }{\partial x_{i}}
\end{align}
where (a) is based on the commutative law and the associative law of $\boxdot$ and $\boxtimes$.

The identity (\ref{Proof: Partial deritivative of Exponential *}) is useful in the proofs of Lemma \ref{LEM: Formula for Directional Derivative} and Lemma \ref{LEM: Derivative of Uc}.
For example, one can deduce from (\ref{Proof: Partial deritivative of Exponential *}) that, $\forall i \in \mathbb{M}\backslash \left\{0\right\}$,
\begin{align}
&\frac{\partial }{\partial x_{i}} H\left(R^{\boxtimes}\left(\mathtt{x}\right)\right) = R'\left(1\right) H\left(\rho^{\boxtimes}\left(\mathtt{x}\right)\boxtimes \frac{\partial \mathtt{x}}{\partial x_{i}}\right) \\
&\frac{\partial }{\partial x_{i}} H\left(\rho^{\boxtimes}\left(\mathtt{x}\right)\right) = H\left(\rho'^{\boxtimes}\left(\mathtt{x}\right)\boxtimes \frac{\partial \mathtt{x}}{\partial x_{i}}\right) \\
&\frac{\partial }{\partial x_{i}} H\left(\mathtt{x} \boxtimes \rho^{\boxtimes}\left(\mathtt{x}\right)\right) = H\left(\rho^{\boxtimes}\left(\mathtt{x}\right)\boxtimes \frac{\partial \mathtt{x}}{\partial x_{i}}\right) + H\left(\mathtt{x} \boxtimes \rho'^{\boxtimes}\left(\mathtt{x}\right)\boxtimes \frac{\partial \mathtt{x}}{\partial x_{i}}\right) \\
&\frac{\partial }{\partial x_{i}} H\left(\mathtt{c}_{\epsilon}\boxdot L^{\boxdot}\left(\rho^{\boxtimes}\left(\mathtt{x}\right)\right)\right) = L'\left(1\right) H\left[\mathtt{c}_{\epsilon}\boxdot \lambda^{\boxdot}\left(\rho^{\boxtimes}\left(\mathtt{x}\right)\right) \boxdot \left( \rho'^{\boxtimes}\left(\mathtt{x}\right)\boxtimes \frac{\partial \mathtt{x}}{\partial x_{i}}\right)\right].
\end{align}

Putting the above together, we have, $\forall i \in \mathbb{M}\backslash \left\{0\right\}$,
\begin{align}
\label{Definition: PotFun for UDE}
\frac{\partial }{\partial x_{i}} U_{\text{s}}\left(\mathtt{x}, \epsilon\right) =& L'\left(1\right) \left\{ H\left(\rho'^{\boxtimes}\left(\mathtt{x}\right)\boxtimes \frac{\partial \mathtt{x}}{\partial x_{i}}\right) - H\left(\mathtt{x} \boxtimes \rho'^{\boxtimes}\left(\mathtt{x}\right)\boxtimes \frac{\partial \mathtt{x}}{\partial x_{i}}\right) \right. \quad\quad\quad\quad\quad\quad\quad \nonumber \\
&\quad\quad\quad- \left. H\left[\mathtt{c}_{\epsilon}\boxdot \lambda^{\boxdot}\left(\rho^{\boxtimes}\left(\mathtt{x}\right)\right) \boxdot \left(\rho'^{\boxtimes}\left(\mathtt{x}\right)\boxtimes \frac{\partial \mathtt{x}}{\partial x_{i}}\right)\right] \right\}
\end{align}
\begin{align}
\stackrel{(a)}{=}& L'\left(1\right) \left\{ H\left[\mathtt{x} \boxdot \left(\rho'^{\boxtimes}\left(\mathtt{x}\right)\boxtimes \frac{\partial \mathtt{x}}{\partial x_{i}}\right)\right] - H\left[\mathtt{c}_{\epsilon}\boxdot \lambda^{\boxdot}\left(\rho^{\boxtimes}\left(\mathtt{x}\right)\right) \boxdot \left(\rho'^{\boxtimes}\left(\mathtt{x}\right)\boxtimes \frac{\partial \mathtt{x}}{\partial x_{i}}\right)\right] \right\} \\
=& L'\left(1\right) H\left[\left(\mathtt{x} - \mathtt{T}_\text{s}^{\left(1\right)}\left(\mathtt{x}, \mathtt{c}_{\epsilon}\right)\right) \boxdot \left(\rho'^{\boxtimes}\left(\mathtt{x}\right)\boxtimes \frac{\partial \mathtt{x}}{\partial x_{i}}\right)\right] \\
\stackrel{(b)}{=}& L'\left(1\right) H\left[\left(\mathtt{T}_\text{s}^{\left(1\right)}\left(\mathtt{x}, \mathtt{c}_{\epsilon}\right) - \mathtt{x}\right) \boxtimes \left(\rho'^{\boxtimes}\left(\mathtt{x}\right)\boxtimes \frac{\partial \mathtt{x}}{\partial x_{i}}\right)\right]
\end{align}
where (a) and (b) follow from (\ref{Corollary: Duality Rule for Entropy Ext1}) and (\ref{Corollary: Duality Rule for Entropy Ext2}), respectively.

By the continuity of $U_{\text{s}}\left(\mathtt{x}, \epsilon\right)$ with respect to $\mathtt{x}$, we have
\begin{align}
\text{d}_{\mathtt{x}}U_{\text{s}}\left(\mathtt{x}, \epsilon\right)\left[\delta\mathtt{x}\right] =& \sum_{i = 1}^{m} \frac{\partial }{\partial x_{i}} U_{\text{s}}\left(\mathtt{x}, \epsilon\right)  \delta x_i \\
=& L'\left(1\right) H\bigg[\left(\mathtt{T}_\text{s}^{\left(1\right)}\left(\mathtt{x}, \mathtt{c}_{\epsilon}\right) - \mathtt{x}\right) \boxtimes \rho'^{\boxtimes}\left(\mathtt{x}\right)\boxtimes \sum_{i = 1}^{m}\frac{\partial \mathtt{x}}{\partial x_{i}} \delta x_i \bigg] \\
\stackrel{(a)}{=}& L'\left(1\right) H\bigg[\left(\mathtt{T}_\text{s}^{\left(1\right)}\left(\mathtt{x}, \mathtt{c}_{\epsilon}\right) - \mathtt{x}\right) \boxtimes \rho'^{\boxtimes}\left(\mathtt{x}\right) \boxtimes \sum_{i = 1}^{m} \left(\mathtt{\Delta}_{i} - \mathtt{\Delta}_{0}\right) \delta x_i \bigg]\\
\stackrel{(b)}{=}& L'\left(1\right) H\left(\left(\mathtt{T}_\text{s}^{\left(1\right)}\left(\mathtt{x}, \mathtt{c}_{\epsilon}\right) - \mathtt{x}\right)\boxtimes \rho'^{\boxtimes}\left(\mathtt{x}\right)\boxtimes\delta\mathtt{x}\right)
\end{align}
where (a) is based on (\ref{Proof: Derivative of x with xi}) and (b) follows from (\ref{Definition: Density Direction 2}).

\section*{Acknowledgment}

The authors would like to thank the reviewers for their insightful comments and thoughtful suggestions on the previous versions of the manuscript.

\ifCLASSOPTIONcaptionsoff
  \newpage
\fi

\bibliographystyle{IEEEtran}
\bibliography{ZZH}

\end{document}